\documentclass[acmsmall]{acmart}

\AtBeginDocument{%
  \providecommand\BibTeX{{%
    \normalfont B\kern-0.5em{\scshape i\kern-0.25em b}\kern-0.8em\TeX}}}

\setcopyright{acmcopyright}
\copyrightyear{2018}
\acmYear{2018}
\acmDOI{XXXXXXX.XXXXXXX}

\acmJournal{TOSEM}
\acmVolume{37}
\acmNumber{4}
\acmArticle{111}
\acmMonth{8}

\usepackage[ruled]{algorithm2e}

\usepackage{verbatim}
\usepackage{times}
\usepackage{multirow}
\usepackage{tabularx}
\usepackage{color} 
\usepackage{graphicx}
\usepackage{booktabs}
\usepackage{subcaption}
\usepackage{times}
\usepackage{makecell}
\usepackage{collectbox}
\usepackage{mdframed}
\usepackage{xcolor}
\usepackage{array}

\SetAlFnt{\small}
\SetAlCapFnt{\small}
\SetAlCapNameFnt{\small}
\SetAlCapHSkip{0pt}
\IncMargin{-\parindent}

\usepackage{listings}

\renewcommand{\arraystretch}{1.2}
\begin{document}

\title{ActivFORMS: A Formally-Founded Model-Based Approach to Engineer Self-Adaptive Systems}

\author{Danny Weyns}
\email{danny.weyns@kuleuven.be}
\orcid{0000-0002-1162-0817}
\affiliation{%
	\institution{Katholieke Universiteit Leuven, Linnaeus University}
	\streetaddress{Celestijnenlaan 200A}
	\city{Leuven}
	\country{Belgium}
	\postcode{3001}
}

\author{Usman M. Iftikhar}
\email{musmanir@gmail.com}
\orcid{0000-0002-1343-5834}
\affiliation{%
	\institution{Katholieke Universiteit Leuven, Linnaeus University}
	\streetaddress{Celestijnenlaan 200A}
	\city{Leuven}
	\country{Belgium}
	\postcode{3001}
}

\renewcommand{\shortauthors}{Weyns and Iftikhar}

\begin{abstract}
Self-adaptation equips a computing system with a feedback loop that enables it dealing with change caused by uncertainties during operation, such as changing availability of resources and fluctuating workloads. To ensure that the system complies with the adaptation goals, recent research suggests the use of formal techniques at runtime. Yet, existing approaches have three limitations that affect their practical applicability: (i) they ignore correctness of the behavior of the feedback loop, (ii) they rely on exhaustive verification at runtime to select adaptation options to realize the adaptation goals, which is time and resource demanding, and (iii) they provide limited or no support for changing adaptation goals at runtime. To tackle these shortcomings, we present ActivFORMS (Active FORmal Models for Self-adaptation). ActivFORMS contributes an end-to-end approach for engineering self-adaptive systems, spanning four main stages of the life cycle of a feedback loop: design, deployment, runtime adaptation, and evolution. We also present ActivFORMS-ta, a tool-supported instance of ActivFORMS that leverages timed automata models and statistical model checking at runtime. We validate the research results using an IoT application for building security monitoring that is deployed in Leuven. The experimental results demonstrate that ActivFORMS supports correctness of the behavior of the feedback loop, achieves the adaptation goals in an efficient way, and supports changing adaptation goals at runtime.
\end{abstract}



\maketitle

\section{Introduction}

Dealing with change has been a challenge for software engineers since the inception of software systems. Technological improvements, dynamics in the market, changing demands, and reported bugs require continuous evolution of systems. Our focus is on changes caused by uncertainties that are difficult to predict before system deployment, such dynamics in the availability of resources and changes of workload over time. As it is often too costly or inefficient to anticipate such uncertainties before deployment, and systems may need to be operational 24/7, the uncertainties need to be dealt with at runtime when the missing knowledge becomes available to resolve the uncertainties. One prominent approach to tackle this problem is \textit{self-adaptation}~\cite{Oreizy1998,Kephart2003,Garlan2004,Kramer2007,Cheng2009,Lemos2013,Weyns2019}.  

The basic idea of self-adaptation is to enhance a software system with a feedback loop that monitors the system and its environment to resolve uncertainties, reasons about the changing conditions, and adapts the system in order to achieve or maintain particular quality requirements (i.e., adaptation goals), or degrades gracefully otherwise. A classic example of a self-adaptive system is an elastic Cloud platform that monitors client applications and automatically adjusts capacity to maintain steady performance at the lowest possible cost. In this research, we focus on \textit{architecture-based adaptation} that is widely considered as an effective approach to cope with uncertainties at runtime~\cite{Oreizy1998,Garlan2004,Kramer2007,Weyns2012-1,Camara16}. On the one hand, architecture-based adaptation offers design abstractions that enable designers to define self-adaptive systems. On the other hand, it provides runtime modeling abstractions that enable a system to reason about change and make effective adaptation decisions~\cite{WeynsBook2020}.

One of the main challenges in engineering self-adaptive systems is providing guarantees that the adaptation goals are met~\cite{Camara2013,Cheng2014,Lemos2017,WeynsBook2020}. Given that uncertainties need to be resolved during operation, it is hard to deliver these guarantees completely before the system is deployed. A variety of approaches have been proposed to provide such guarantees, ranging from formal proof to runtime testing~\cite{Weyns2012-2,Tamura2014,Cheng2014,Weyns:2016}. Our focus here is on the use of formal modeling and verification techniques, the most popular approach~\cite{Lemos2017}. An example is a service-based system equipped with a feedback loop that maintains a parameterized Markov model of the system and quality goals that are expressed as probabilistic logic formulae, enabling runtime model checking to identify optimal system configurations~\cite{Calinescu2011}. Yet, existing approaches have three limitations that affect their practical applicability: (i) they ignore correctness of the behavior of the feedback loop, (ii) they rely on exhaustive verification at runtime to select adaptation options to realize the adaptation goals, which suffers from the state explosion problem~\cite{Clarke2008}, and  (iii) they provide limited or no support for changing adaptation goals at runtime. Overall, existing approaches lack a comprehensive perspective on the adaptation problem~\cite{2013-SEFSAS2b}. 

To tackle these limitations, this paper presents ActivFORMS (Active FORmal Models for Self-adaptation) with the following first contribution:  

\begin{itemize}
	
	\item[]\textit{ActivFORMS: a reusable end-to-end model-driven approach for engineering self-adaptive 
	systems that spans four main stages of the life cycle of a feedback loop: design, deployment, runtime adaptation, and evolution.}
	\vspace{5pt}
\end{itemize}

\noindent ActivFORMS supports the different stages of the life cycle of a feedback loop as follows: 

\begin{enumerate}
	
	\item At design time, a feedback loop is specified using formal models. 
	ActivFORMS provides guarantees for the correct behavior of the feedback loop with respect to the set of correctness properties through model checking of the feedback loop models.
	
	\item At deployment time, the verified feedback loop models are directly deployed and then executed to realize adaptation using a model execution engine. As such, ActivFORMS preserves the guarantees obtained at design through direct execution of the verified models.
	
	\item At runtime, the feedback loop selects adaptation options that realize the adaptation goals in an efficient manner. ActivFORMS guides the adaptation of the system at runtime by analyzing and selecting adaptation options that realize the adaptation goals with a required level of accuracy and confidence in an efficient manner using statistical model verification techniques at runtime.
	
	\item At evolution time, ActivFORMS offers basic support for online evolution of the feedback loop. In particular, ActivFORMS offers basic support for changing adaptation goals and updating the verified models of the feedback loop on-the-fly to meet the new goals.
	
\end{enumerate}

The reusability of ActivFORMS lays in its flexibility to define different instances of the approach for different types of self-adaptive systems.
This paper presents such an instance called ActivFORMS-ta (``ta'' for ``timed automata''), with the following second contribution: 

\begin{itemize}
	\item[]\textit{ActivFORMS-ta: a tool-supported instance of ActivFORMS that: (i) supports the design of feedback loop models as networks of timed automata and the verification of their correctness, (ii) enables direct execution of the models, (iii) realizes the adaptation goals in an efficient manner, (iv) supports adding new goals and dynamic updates of feedback loop models.}
	\vspace{5pt}
	
\end{itemize}

ActivFORMS-ta offers concrete artifacts that were defined only once and from then can be reused for realizing self-adaptation for a broad family of systems. These artifacts include (i) a set of \textit{model templates} that supports the design and verification of feedback loop models, (ii) a \textit{trusted virtual machine} that directly executes the verified feedback loop models to realize adaptation, and (iii) a \textit{trusted online update manager} that can be used to update goals and feedback loop models on-the-fly.  

In line with current research, we focus on runtime uncertainties related to parameters of the system or the environment~\cite{Malek2011,Perez-Palacin:2014:UMS:2568088.2568095,MAHDAVIHEZAVEHI201745,9196226}. We also consider uncertainties related to the feedback loop functions and adaptation goals, to the extent that they can be handled by online updates of the feedback loop models. Uncertainties that require evolution of the managed system are out of scope.

We validate ActivFORMS and its instance ActivFORMS-ta using a real-world Internet-of-Things (IoT) application for monitoring buildings. We compare the results with a state-of-the-art approach and a conservative approach based on current practice. The validation demonstrates the instantiation of ActivFORMS and the contributions of ActivFORMS-ta for a practical self-adaptive system.

The remainder of this paper is structured as follows. In Section~\ref{section:preliminaries} we give background on architecture-based adaptation, timed automata, and statistical model checking, and we introduce the IoT application. Section~\ref{section:overview} gives a high-level overview of ActivFORMS with its four stages. Sections~\ref{section:stage_I} to \ref{section:stage_IV} zoom in on these stages one by one. In Section~\ref{section:evaluation}, we present the evaluation results. Section~\ref{section:related-work} discusses related approaches and summarizes initial efforts on which this paper leverages. Finally, we draw conclusions and outline opportunities for future research in Section~\ref{section:conclusions}.

\section{Preliminaries}\label{section:preliminaries}
We provide background on architecture-based adaptation, timed automata, and statistical model checking, introducing basic terminology and concepts. Then, we introduce the IoT application. 

\subsection{Architecture-Based Adaptation and MAPE-K Feedback Loop}\label{section:architecture-based-adaptation}

\begin{figure}[h!tb]
	\centering
	\includegraphics[width=0.9\textwidth]{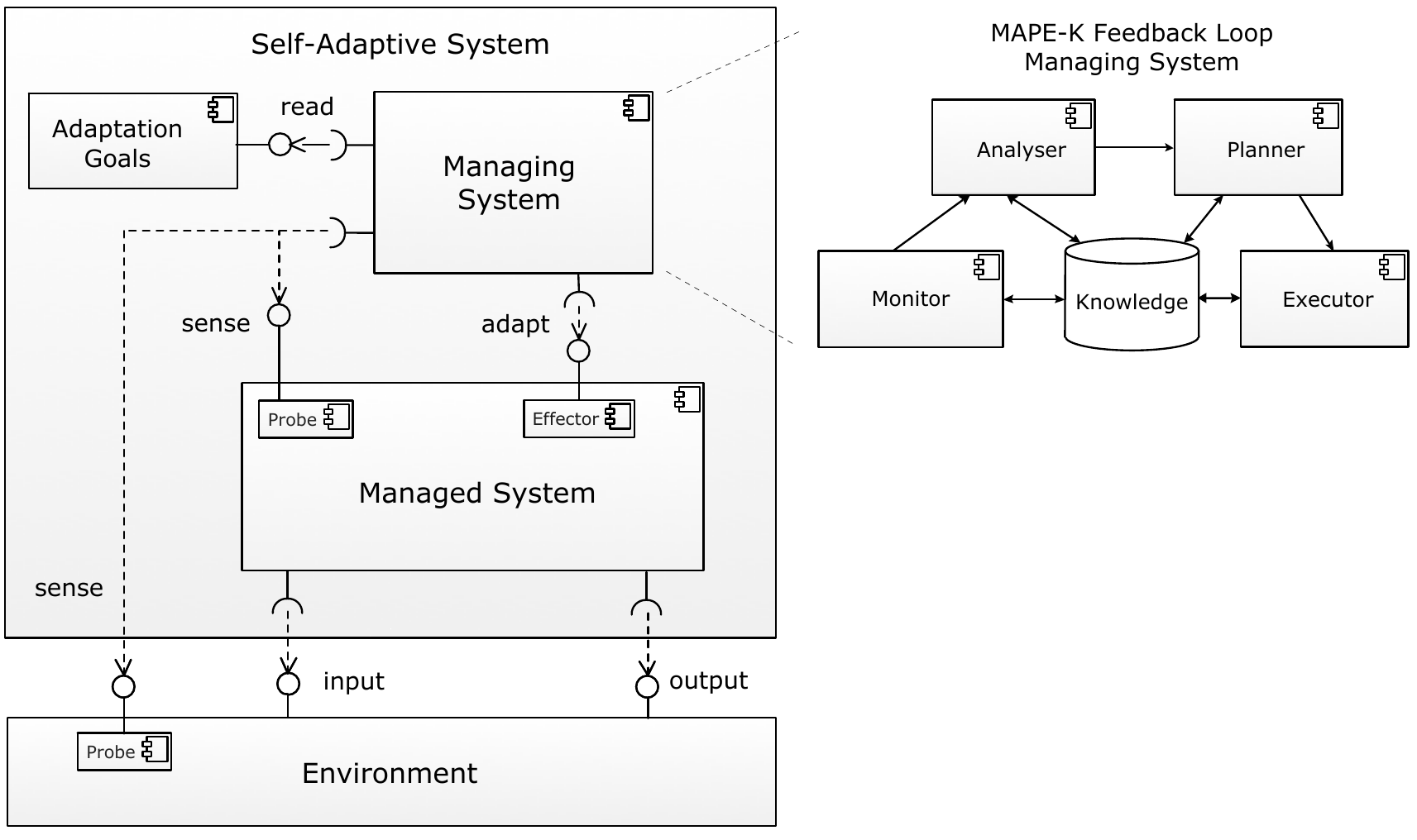}
	\caption{Basic building blocks of a self-adaptive system with MAPKE-K feedback loop}
	\label{fig:SAS}
\end{figure}

The key driver for self-adaptation is enabling a software system to deal with uncertainties that are hard or impossible to anticipate before deployment, such as changing availability of resources and fluctuating work loads. If not treated properly, such uncertainties can jeopardize the system's quality goals (performance, reliability, etc.). The idea of self-adaptation is to let the system collect additional data when they become available to resolve uncertainties and adapt the system accordingly such that it maintains its quality goals, or degrade gracefully if necessary. 
Despite two decades of active research~\cite{Oreizy1998,Kephart2003,Garlan2004,Kramer2007,Cheng2009,Lemos2013,Lemos2017,WeynsBook2020}, there is no commonly agreed definition of self-adaptation.  However, there are two common ways to look at self-adaptation~\cite{WeynsBook2020}: (1) the ability of a system to adjust its behavior in response to changes of the environment and the system~\cite{Cheng2009}; the ``self'' prefix indicates that the system adapts autonomously (or with minimal human intervention)~\cite{Brun:2009}, and (2) the mechanisms that are used to realize self-adaptation, typically a closed feedback loop~\cite{Dobson2006,Salehie:2009}. Fig.~\ref{fig:SAS} shows the basic building blocks of a self-adaptive system that integrates these two views~\cite{WeynsBook2020}. 

The \textit{Environment} refers to the part of the external world with which the self-adaptive system interacts and in which the effects of the system can be observed~\cite{1997Jackson}. The \textit{Managed System} comprises the application that realizes the system's domain functionality. The \textit{Managing System} comprises the adaptation logic that deals with one or more adaptation goals. The managing system can \textit{sense} the managed system and its environment through a \textit{Probe}, and it can \textit{adapt} the managed system through an \textit{Effector}. The \textit{Adaptation Goals} are concerns about the managed system that are dealt with by the managing system; they usually relate to software qualities of the managed system~\cite{6224395}. The adaptation goals can be subject to change themselves (which is not shown in Fig.~\ref{fig:SAS}). 

A typical approach to structure the managing system is by means of a MAPE-K feedback loop (Monitor-Analyzer-Planner-Executer elements that share Knowledge~\cite{Kephart2003}). The \textit{Monitor} collects runtime data from the managed system and the environment and uses this to update the \textit{Knowledge}. This runtime data helps resolving uncertainties. The \textit{Analyzer} uses the knowledge to determine whether there is a need for adaptation of the managed system, and if so it analyses the options for adaptation. The \textit{Planner} then selects the best option and composes a plan\footnote{We use ``Planner'' as the common name of the module that determines the steps that need to be performed to adapt a system.} consisting of a set of adaptation actions that are then enacted by the \textit{Executor} that adapts the managed system as needed.

\subsection{Timed Automata}\label{section:timed_automata}

A timed automaton~\cite{ALUR1994183} is a finite state machine extended with a set of real-valued clocks that progress synchronously.
Formally, a timed automaton can be defined as a tuple $\mathit {TA}$\,=\,$(L, l_{0}, C, A, E, I)$:
\begin{itemize}
	\item[] $L$ is a finite set of locations or states;
	\item[] $l_{0} \in L$ is the initial location of the automaton;
	\item[] $C$ is a finite set of clocks that can be reset;
	\item[] $A$ is a finite set of actions that include synchronization actions and internal actions;
	\item[] $E$\,\,$\subseteq$\,\,$L$\,$\times$\,$A$\,$\times$\,$B(C)$\,$\times$\,$2^C$\,$\times$\,$L$ is a set of edges that connect locations with an action, a guard, a set of clocks. $B(C)$ represents clock constraints over edges and locations;
	\item[] $I$\,:\,$L$$\rightarrow$$B(C)$ assigns invariants to locations.
\end{itemize}

Automata (or behaviors) can be connected in a network. The state of the system is then defined by the state of all automata, the clock values, and the values of the ordinary variables. Only one state  per automaton, called \emph{control} or \emph{active} state (or current location), is active at a time. Automata can synchronize through channels, where a sender $x!$ can synchronize with a receiver $x?$ through a signal $x$. The sender will be blocked if there is no receiver.\footnote{Automata can also synchronize through a (non-blocking) broadcast channel where a sender sends a signal to a set of receivers.} An edge can be annotated with: a \textit{guard}, i.e., a condition on the values of clocks and variables that must be satisfied to take the edge (e.g., $y<5$); a \textit{synchronization} 
action (e.g., $x!$) that forces a synchronization with a complementary action (e.g., $x?$) when the edge is taken; and an \textit{update} action to be taken when a transition is made (e.g., a function $reset()$ resets
clock $y$ to 0). The absence of a guard corresponds to a condition $true$.

Uppaal~\cite{tutorial04} offers a suite that supports modeling and verification of networks of automata. Graphical specifications can be complemented with expressions specified in a C-like language to define data structures (\textit{struct} concept) and functions.\footnote{For the Uppaal grammar, see https://www.uppaal.org/} Goals can be expressed in Timed Computation Tree Logic (TCTL). TCTL expressions describe state and path formulae that can be verified, such as reachability (a system should/can/cannot/... reach particular states), liveness (something eventually will hold), etc. Uppaal defines two types of transitions between states: \emph{action transition} and \emph{delayed transition}. Action transitions can be further divided into \emph{synchronization transition} and \emph{internal transition}. In a synchronization transition, automata synchronize via a channel as explained above. In an internal transition, an automata moves from its current state (say $\mathit{S}$) to a next state  ($\mathit{T}$) via an edge ($e$) when the conditions hold to make the transition, e.g., the guard on the edge is satisfied, the invariant of $\mathit{T}$ holds, etc.  
In a delayed transition only the clocks tick and no actual state transition is made (e.g., $\mathit{S}$ remains active while an invariant such as $y < $ \textit{MAX\_TIME} holds). Further progress in time might lead to an invariant violation ($y \geq$  \textit{MAX\_TIME}) triggering a transition ($\mathit{S}$$\rightarrow$$\mathit{T}$). 

To enable modeling of atomicity of transition sequences (i.e., multiple transitions with no time delay) states
may be marked as committed (marked with a C). If a control state of one of the behaviors is labeled committed, no delay is allowed to occur and any action transition (synchronization or not) must involve the particular behavior (the behavior is so to speak committed to continue)~\cite{Larsen97uppaalin}. The semantics of urgent states (marked with a U) is the same as: introducing a new clock; reset the new clock on all in-going transitions to the location; and add a conjunct to the location invariant requiring the new clock to be 0. Intuitively, this forces the process to leave an urgent location without delay.\footnote{See the Uppaal help pages at https://www.uppaal.org/}

To support stochastic behaviours, a stochastic interpretation of the timed automata has been proposed that replaces the non-deterministic choices between multiple enabled transitions by probabilistic choices. Similarly, non-deterministic choices of time delays are refined by probability distributions, enabling to include quantitative information about the likelihood of choice alternatives. 

Uppaal-SMC~\cite{David2015} supports modelling and verification of networks of stochastic timed automata. Here it is assumed that the automata are input-enabled, determistic (based on a probability measure), and non-zeno.\footnote{5A system behavior is called zeno if it includes an infinite number of discrete steps in a finite amount of time. A specification that is not zeno is said to be non-zeno.} The communication in these networks is restricted to broadcast synchronizations to keep a clean semantics of only non-blocked components which are racing against each other with their corresponding local distributions. Uppaal-SMC supports discrete probabilistic choices based on weights and applies independent,
uniform distributions as default probabilistic choices.\footnote{Uppaal-SMC also supports exponential distributions with user-supplied rates for states without invariants, where an automaton can remain indefinitely. Uppaal-SMC also supports discrete probabilistic choices based on weights.}
In this paper, all models rely either on default uniform distributions of probabilistic choices for bounded delays, or discrete probabilistic choices based on weights.

\subsection{Statistical Model Checking}\label{section:statistical_model_checking}

Statistical model checking (SMC) offers an efficient alternative to traditional model checking that exhaustively traverses all the states of the system, see for instance~\cite{Legay2010,Agha:2018}.  
The central idea of SMC is to check the probability $p\in[0,1]$ that a hypothesized model $M$ of a stochastic system satisfies a property $\varphi$, i.e., to check $P_{M}(\varphi)\geq p$ by performing a series of simulations. SMC applies statistical techniques on the simulation results to decide whether the system satisfies the property with some degree of accuracy and confidence. 
Two types of statistical inference are applied: (a) \textit{hypothesis testing} determines the extent to which observations conform to a given specification; and (b) \textit{estimation} determines likely values of parameters based on the assumption that the data is randomly drawn from a specified type of distribution. The results of an inference can be used to evaluate a property specified in a stochastic temporal logic~\cite{Agha:2018}. 

Uppaal-SMC~\cite{David2015} is a tool for statistical model checking that supports different types of verification queries; here we focus on probability estimation (hypothesis testing) and simulation (estimation). Probability estimation computes an estimation of probability $p$ for an expression $\varphi$ with an approximation interval $[p-\epsilon, p+\epsilon]$ and confidence $1-\alpha$ in a given time $bound$.\footnote{To determine apriori the number of (simulation) runs that are required based on the values of $\epsilon$ and $\alpha$, Chernoff-Hoeffding inequality can be used \cite{Hoeffding}. Uppaal-SMC implements a more efficient sequential method where a probability confidence interval (for given $\alpha$) is derived with each new simulation measurement and the simulation generation is stopped when the confidence interval width is less than 2$\epsilon$. For further details, we refer the interested reader to \cite{David2015}.} The approximation interval provides a range of values which is likely to contain $p$ with a confidence level selected by the user; e.g., the approximation interval may provide estimations of the probability $p$ within an interval of $\pm$~1\% with a confidence level of 95~\%. A confidence stated as $1$\,$-$\,$\alpha$ can be thought of as the inverse of a significance level $\alpha$. A probability estimation query is formulated as $p$\,$=$\,$Pr[bound](\varphi)$. The second type of query, simulation, performs $N$ simulation runs of the system model in a time $bound$ to provide insight in the values of expected system behaviors.
A simulation query is formulated as $simulate~N [\leq bound]\{E1,...,Ek\}$, where $N$ is a natural number indicating the number of simulations to be performed, $bound$ is the time bound on the simulations, and $E1, ..., Ek$ are the (state-based) expressions that need to be monitored during the simulation.\footnote{Note that the modeler can use both $N$ and $bound$ as a means to determine how the simulation runs are executed.} A simulation query can for instance be used to determine the expected value of a variable of a model for a series of simulation runs. 

A benefit of SMC is that the parameters $\alpha$, $\epsilon$, $N$ and $bound$ allow designers to tradeoff the accuracy and confidence of the results with the resources and the time required for verification. For probability estimation, more accurate results (lower $\epsilon$) and higher confidence (lower $\alpha$) require more resources and verification time, and vice versa. Similarly, for a simulation query, more simulation runs (higher $N$) provide more accurate results. When determining the parameter settings of verification queries, the designer should take into account that in general the number of simulation runs is polynomial in 1/$\epsilon$ and log 1/$\alpha$ \cite{Herault2004}. It is important to note that in contrast to exhaustive verification approaches (such as runtime quantitative verification \cite{Calinescu2011}), a simulation-based approach does not provide 100\% guarantees, but an estimation that is bound to an accuracy interval and level of confidence  \cite{Younes2004,Clarke2008,David2015}.  

In this research, we do not consider rare events for which specific techniques such as importance sampling and importance splitting may be applied to statistical model checking \cite{Legay2016}.

\subsection{Self-Adaptive Internet-of-Things Application}\label{section:iot}

We now introduce an IoT application, called DeltaIoT, that we use to illustrate ActivFORMS and its evaluation.
DeltaIoT is a reference IoT application that provides the characteristics (uncertainties, conflicting adaptation goals, distribution, etc.) of a challenging case for applying self-adaptation~\cite{Iftikhar2017} and has been used frequently, see e.g.,~\cite{RESTREPO2021111010}. The network offers both a physical setup for field experimentations and a simulator for offline experimentations. DeltaIoT is part of the smart campus initiative\footnote{https://people.cs.kuleuven.be/danny.weyns/software/DeltaIoT/} by DistriNet, KULeuven in collaboration with VersaSense, a provider of industrial IoT.\footnote{https://versasense.com/} 

\begin{figure}[h!tb]
	\centering
	\includegraphics[width=0.95\textwidth]{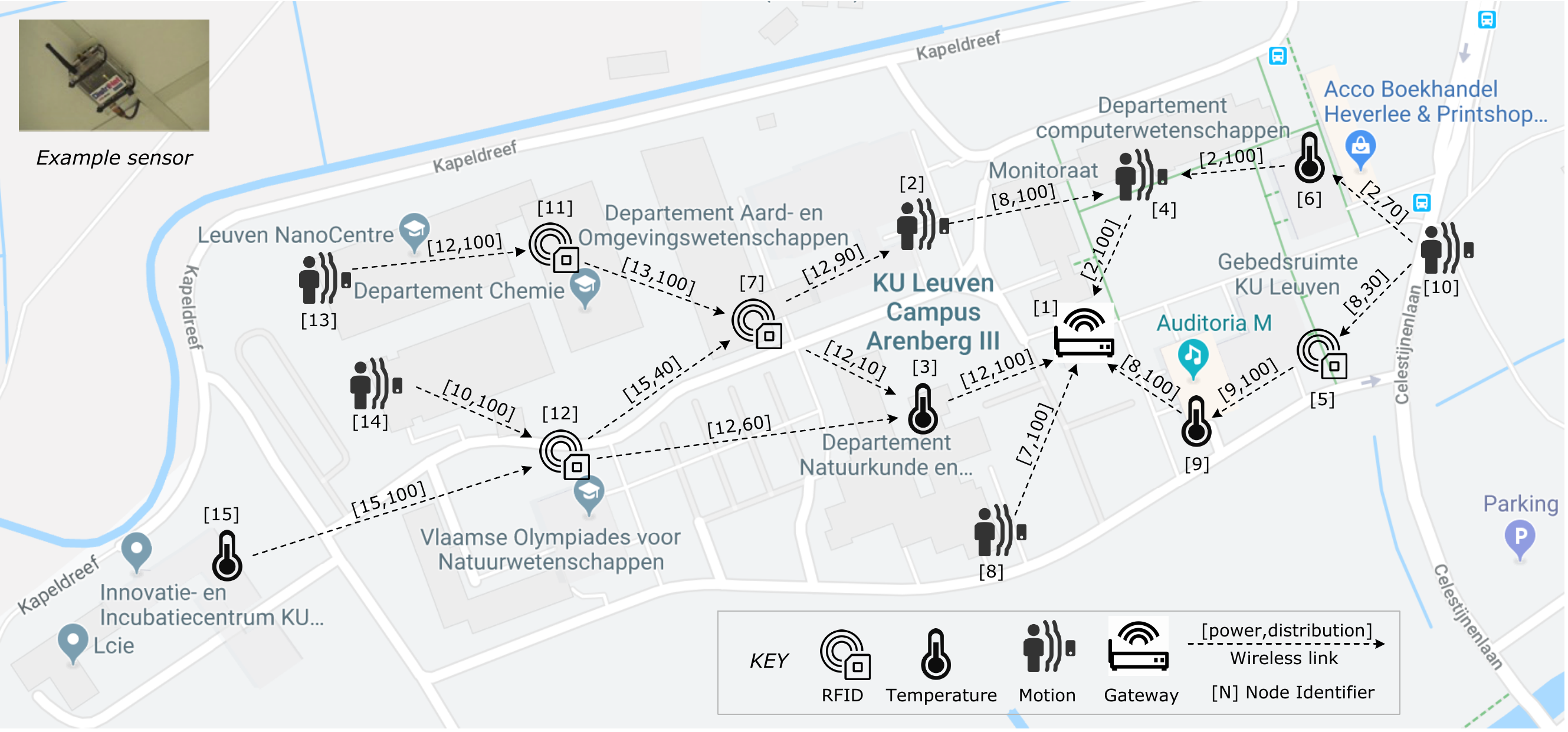}
	\caption{DeltaIoT system with network topology and example of a sensor}
	\label{fig:DeltaIoT}
\end{figure}

DeltaIoT consists of a collection of 15 battery-powered LoRa-based\footnote{https://www.lora-alliance.org/What-Is-LoRa/Technology} IoT motes that are deployed at the KULeuven campus, see Fig.~\ref{fig:DeltaIoT}. In each building, motes are strategically placed to provide access control to labs (RFID sensor), to monitor the occupancy status (passive infrared sensor) and to sense the temperature (heat sensor, an example is show top right of Fig.~\ref{fig:DeltaIoT}). The sensor data from all the motes are relayed to the IoT gateway, which is deployed at a central monitoring facility. Campus security staff can monitor the status of buildings and labs from the monitoring facility and take appropriate action whenever unusual behavior is detected in the buildings.

DeltaIoT uses multi-hop wireless communication. As shown in Fig.~\ref{fig:DeltaIoT}, each IoT mote in the network relays its sensor data to the gateway, either directly or via intermediate IoT motes.\footnote{A sending mote is a child from the viewpoint of a receiving mote and the receiving mote is then parent of the sending mote.} DeltaIoT uses time synchronized communication~\cite{Dujovne2014}. Concretely, the communication in the network is organized in cycles, each cycle comprising a fixed number of communication slots. Each slot defines a sender mote and a receiver mote that can communicate with one another. The communication slots are fairly divided among the motes. For example, the system can be configured with a cycle time of 570 second (9.5 minutes) with each cycle comprising 285 slots, each of 2 seconds. For each link, 40 slots are allocated for communication between the motes.

Each mote is equipped with three queues: \textit{buffer} collects the packets produced by the mote, \textit{receive-queue} collects the packets from the mote's children, and \textit{send-queue} queues the packets to be sent to the parent(s) during the next cycle. The size of the \textit{send-queue} is equal to the number of slots that are allocated to the mote for communication during one cycle. Before communicating, the packets of the \textit{buffer} are first moved to the \textit{send-queue}; the remaining space is then filled with packets from the \textit{receive-queue}. Packets that arrive when the \textit{receive-queue} is full are lost (i.e., \textit{queue loss}). 

IoT applications are expected to last a long time on a set of batteries (typically multiple years), while offering reliable communication with minimal latency. To guarantee these quality properties, the motes of the network should be optimally configured. Two key factors that determine the critical quality properties are the transmission power of the motes and the selection of paths to relay packets towards the gateway (i.e., the distribution of the packets sent via the links to the respective parents). Guaranteeing the required quality properties is complex as the system is subject to various types of uncertainties. Here, we consider two primary types of uncertainty: 

\begin{enumerate}
	\item \textit{Network interference and noise:} Due to external factors such as weather conditions and the presence of wireless signals such as WiFi in the neighborhood, the quality of the communication between motes may be affected, which in turn may lead to packet loss.
	\item \textit{Fluctuating traffic load:} The packets produced by the motes may fluctuate in ways that are difficult to predict (e.g., packets produced by a passive infrared sensor are based on the detection of motion of humans).
\end{enumerate}

\begin{figure}[h!tb]
	\centering
	\includegraphics[width=0.75\textwidth]{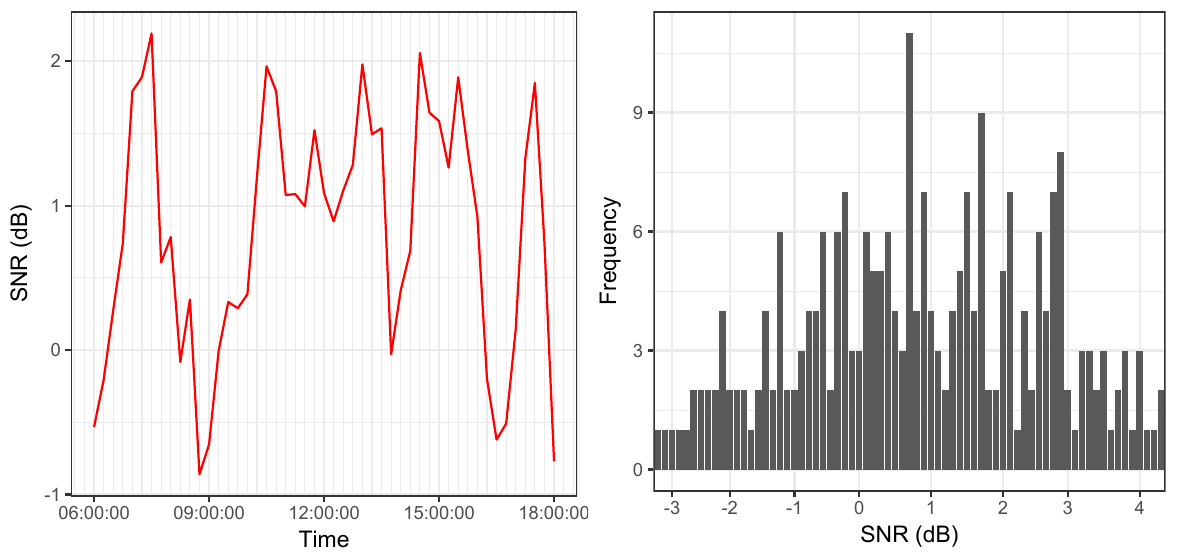}
	\caption{Profiles of uncertainties for one of the motes in Figure~\ref{fig:DeltaIoT}.}
	\label{fig:uncertainty_profiles}
\end{figure}

As an example, the graph on the left of Fig.~\ref{fig:uncertainty_profiles} shows the values of the signal to noise ratio of the communication link between two motes over time. Signal to Noise Ratio ($\mathit{SNR}$ in decibels $dB$) represents the ratio between the levels of desired signal and undesired signal, i.e., noise, which comes from the environment. The higher the interference, the lower the SNR, resulting in higher packet loss. The graph on the right hand side shows the frequency of the same data with a resolution of one digit, which has a normal distribution.\footnote{The graphs in Fig.~\ref{fig:uncertainty_profiles} are based on values collected during field observations for a period of one week. We use these graphs as profiles for the uncertainties in simulation mode. The Shapiro-Wilk test gave a p-value of 0.06; with a significance level 0.05.}

The quality requirements for DeltaIoT that become adaptation goals for self-adaptation are:
\begin{enumerate}
	\item[] $R1$: The average packet loss per period of 12 hours should not exceed 10\%;    
	\item[] $R2$: The energy consumption should be minimized. 
\end{enumerate}

In addition, the following adaptation goal should be added  to the system during operation:   

\begin{enumerate}
	\item[] $R3$: The average latency of packets per 12 hours  should be less than 5\% of the cycle time. 
\end{enumerate}

DeltaIoT also requires the following failsafe operating mode when adaptation is applied: 

\begin{enumerate}
	\item[] $R4$: If no valid adaptation option is available, apply the reference setting; i.e., set the transmission power of motes to maximum and duplicate all packets to all parents. 
\end{enumerate}

\vspace{+2pt}
\noindent
\textbf{Why Self-adaptation?} The key problem of DeltaIoT is how to ensure the quality goals regardless of the uncertainties in network interference and fluctuating traffic load of packets. A typical approach used in practice to deal with the uncertainties in IoT  applications such as DeltaIoT is to over-provision the network. In this approach, the transmission power of the links is set to maximum and all packets transmitted by a mote are copied to all its parents. Operators may fine-tune these settings based on trial-and-error using observations of the network. While such a conservative approach may result in low packet loss, the cost is high energy consumption. Furthermore, manual intervention is a costly and error-prone activity. By enhancing DeltaIoT with self-adaptation capabilities, the system will automatically track the uncertainties at runtime and use up-to-date information to find and adapt the settings of the motes such that the system complies with the quality requirements.

\vspace{+8pt}
\noindent
\textbf{Interface Implementation.} DeltaIoT offers a client deployed at the gateway that includes a Java package with Probe and Effector classes. Listing~\ref{probe} shows the main methods of the probe to monitor the network and the effector to adapt the mote settings (for the physical network and the simulator). 

\lstset{caption={DeltaIoT probe and effector methods.},label=probe}
\footnotesize
{\ttfamily
	\begin{lstlisting}
	ArrayList<Mote> getAllMotes(); 
	ArrayList<QoS> getNetworkQoS(Period);
	void setMoteSettings(MoteID, List<LinkSetting>); 
	void resetDefaultConfiguration();
	\end{lstlisting}
}
\normalsize

$getAllMotes$ returns an array with a representation of each mote of the network for a cycle, including the traffic generated by a mote (number of messages sent from 0 to 10), the  energy consumed (in Coulomb), the settings of the transmission power that a mote used to communicate with each of its parent (in a range from 0 to 15), the spreading factor used for each link (7 to 12),\footnote{Technically, the spreading factor is defined as the number of chirps used per symbol, where the chirp rate is equal to the bandwidth~\cite{Augustin2016}. A higher spreading factor results in longer range, at the cost of more energy consumption.} the SNR for each link (in dB, typically in the range of 10 to -40), and the distribution factor per link being the percentage of the packets sent by a source mote over the link to each of its parents (0 to 100\%).\footnote{The sum of the distribution factors for a mote is 100, but when packets are duplicated to more parents, the sum is above 100.}  $getNetworkQoS$ returns statistical data about the quality of service (QoS) of the overall network for a given period. Currently this method returns data about packet loss (fraction of packets lost [0...1]), energy consumption (Coulomb), and latency of the network (fraction of the cycle time that packets remain in the network [0...1]; 0 means all packets are delivered in the cycle they are generated; 1 means packets all packets are delivered in the cycle after the cycle they are generated). 

$setMoteSettings$ can be used to set the parameters for the parent links of a mote with a given ID. A $LinkSetting$ contains the source and destination node of the link, the transmission power to be used to communicate via the link (0 to 15), and the distribution factor for the link (0 to 100\% in steps of 20\%). Finally, $resetDefaultConfiguration$ resets the network settings to predefined values. This method can be used to bring the system to a well-known state, e.g., as failsafe state. 

\section{High-Level Overview of the ActivFORMS Approach}\label{section:overview}

ActivFORMS offers a reusable approach to engineer self-adaptive software systems. The approach combines: (i) design-time correct-by-construction modeling of the feedback loop, (ii) deployment and direct execution of the verified feedback loop model to realize adaptation, (iii) runtime statistical model checking to infer quality estimates of different system configurations; the estimates are then used to guide the adaptation of the system to realize the adaptation goals, and (iv) basic support for on-the-fly updates of adaptation goals and the feedback loop model when needed.

ActivFORMS supports self-adaptive software systems based on MAPE feedback loops~\cite{Kephart2003,Dobson2006,Calinescu2011,exp}. Other types of feedback loops, e.g., based on principles from control theory (see~\cite{Shevtsov17} for a survey), are not supported by ActivFORMS.

ActivFORMS relies on three basic principles:

\begin{enumerate}
	
	\item Model-driven: models are the central artifacts in ActivFORMS to realize self-adaptation, from design time to operation and evolution, using model-based specification and formal verification, direct model execution, model-based analysis, and dynamic model updates.
	\item Continuous verification: in ActivFORMS evidence for the correct behavior of the feedback loop is generated at design time, before deployment of the feedback loop model or updates of the model, and evidence that adaptation options are selected that guide the adaptation of the managed system to realize the adaptation goals is continuously generated at runtime.
	\item Reuse: reusable model templates for the design and verification of feedback loop models, a reusable model execution engine that enables direct execution of a feedback loop model, and a reusable update manager to guide model evolution reduce the effort to engineer self-adaptive systems with ActivFORMS.  
\end{enumerate}

\begin{figure}[h!tb]
	\centering
	\includegraphics[width=\textwidth]{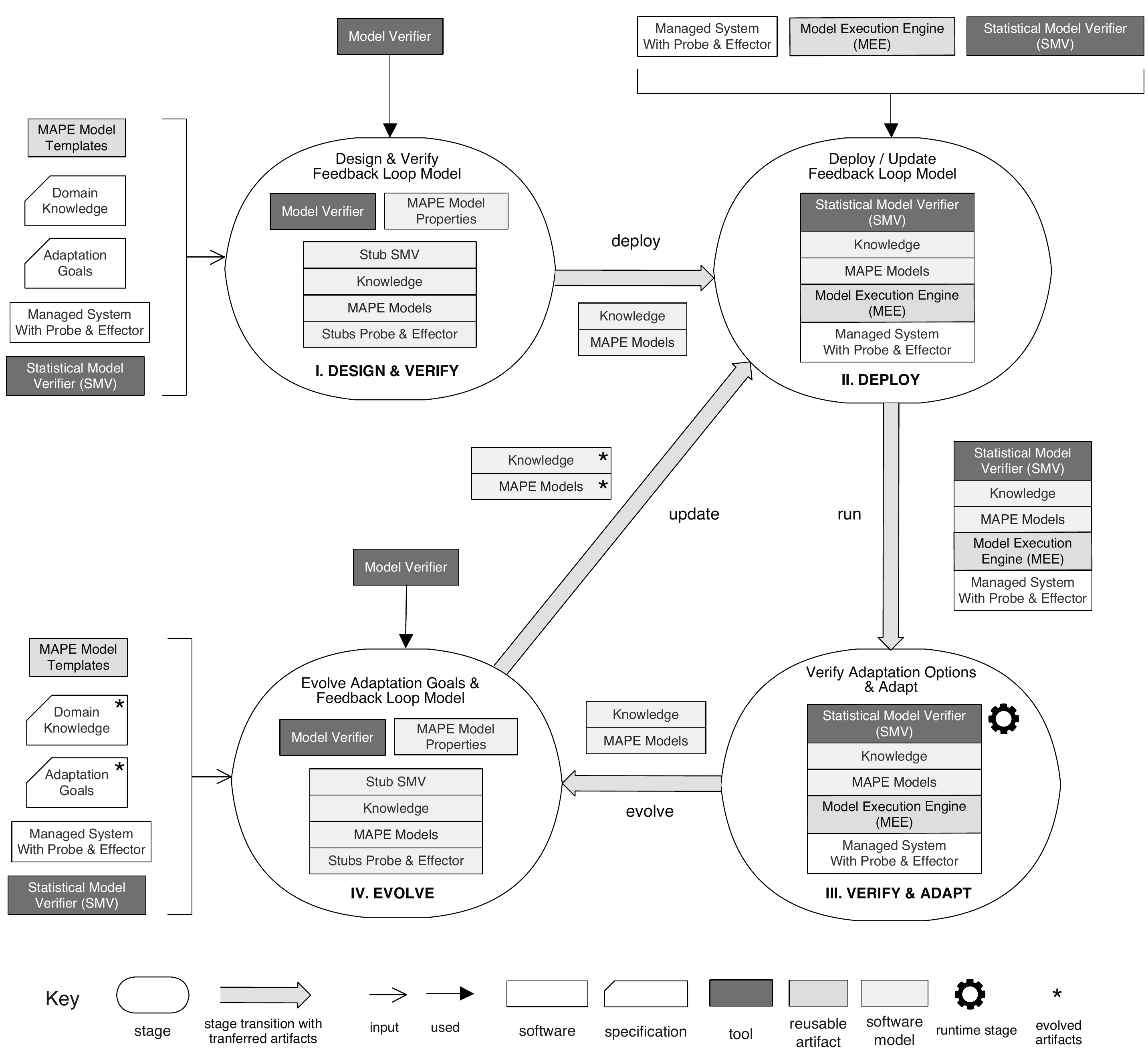}
	\caption{The four stages of ActivFORMS: I. Design \& Verify, II. Deploy, III. Verify \& Adapt, and IV. Evolve}
	\label{fig:activFORMS}
\end{figure}

Fig.~\ref{fig:activFORMS} gives a high-level overview of ActivFORMS that spans four main stages of the software lifecycle of feedback loops. Design \& Deployment, the first two stages, cover the design, verification, and enactment of a feedback loop model. Central to the first two stages are MAPE model templates that support the engineer with the design and verification of a feedback loop model and a model execution engine that executes the verified feedback loop model. Runtime, the third stage, realizes adaptation of the managed system during operation using the deployed feedback loop model to achieve the adaptation goals. Central to the third stage is efficient runtime analysis of first-class quality models relying on statistical model verification. This stage covers ``change management'' in the reference model for self-adaptive systems of~\cite{Kramer2007}. Evolution, the fourth stage, realizes evolution of adaptation goals and the feedback loop model to deal with new or changing goals and updating runtime models. Central to this stage are an online update manager that supports the evolution process. This stage covers ``goal management'' in Kramer and Magee's reference model. 

\textit{ActivFORMS-ta} offers a concrete instance of ActivFORMS.  Table~\ref{instantation-summary} summarizes the instantation. 

\begin{table}[h!]\caption{Summary of instantiation of  ActivFORMS with ActivFORMS-ta}\label{instantation-summary}
	\centering
		\begin{footnotesize}	
		\renewcommand{\arraystretch}{1.2}
		\setlength{\tabcolsep}{0.4em}
		\begin{tabular}{p{3cm}p{9.5cm}}
			\Xhline{0.6pt}
			\textbf{ActivFORMS}   
			&  \textbf{ActivFORMS-ta}
			\\ \Xhline{0.6pt}
			MAPE model templates 
			& MAPE model templates based on timed automata with Uppaal  suite~\cite{Behrmann2004} \\ 	\Xhline{0.4pt}
			Quality models 
			& Stochastic timed automata  
			\\ 	\Xhline{0.4pt}
			Model execution engine 
			& Trusted virtual machine to execute timed automata 
			\\ 	\Xhline{0.4pt}
			Runtime analysis
			& Statistical model checking using Uppaal-SMC~\cite{David2015}
			\\ 	\Xhline{0.4pt}
			Goal management 
			& Trusted online update manager 		`			%
			\\ \Xhline{0.6pt}
		\end{tabular}
	\end{footnotesize}
\end{table}

The main reusable artifacts of ActivFORMS-ta are concrete \textit{MAPE model templates}, a \textit{trusted virtual machine} that directly executes verified MAPE feedback loop models, and a \textit{trusted online update manager} that can be used to update feedback models on-the-fly. With ``trusted'' we mean that evidence is available (obtained through extensive testing) for stakeholders to be confident that the virtual machine and the online update manager will perform their tasks in a reliable way. 
Creating the different artifacts of ActivFORMS-ta took several months. The model templates were designed in iterations and applied until they became stable. This effort took several man-months in total. The design and development of the execution engine and the update manager took also a number of man-months. However, after these initial efforts, these artifacts could be reused with minimal effort. 

The ActivFORMS approach relies on a number of assumptions: (i) the managed system exists and the adaptation goals are known, (ii) probes for monitoring and effectors for adapting the managed system are available, (iii) dynamics in the environment are significantly slower than the execution of adaptations, (iv) the managed system has a limited, possibly high number of configurations (adaptation options) that can dynamically change over time (system parameters with a continuous domain should be discretizable), (v) designers have access to the necessary domain knowledge of the managed system and its environment to design the feedback loop. 

All the artifacts of ActivFORMS-ta, including the MAPE model templates, the virtual machine and the online update manager together with comprehensive test suites and complete test reports, as well as all the material and test results of the concrete application of ActivFORMS to DeltaIoT is available at the ActivFORMS website.\footnote{ActivFORMS website: https://people.cs.kuleuven.be/danny.weyns/software/ActivFORMS/}

In the next four sections, we explain the four stages of ActivFORMS and their instantiation for ActivFORMS-ta in detail. Each stage is illustrated with examples of DeltaIoT. We focus on the main parts and point to the ActivFORMS website for additional parts.

\section{Stage I ActivFORMS: Design and Verify Feedback Loop Model}\label{section:stage_I}

In Stage I of ActivFORMS, a formally verified feedback loop model of the self-adaptive system is created, see Fig.~\ref{fig:activFORMS}. This includes the design of a feedback loop model and its verification. For each activity, we explain the general principles of ActivFORMS, we apply them to ActivFORMS-ta, and we illustrate them for DeltaIoT. We use the same structure for the other stages in the next sections.

\subsection{Design Feedback Loop Model}

\subsubsection{Design Feedback Loop Model with ActivFORMS}

In ActivFORMS, a feedback loop is realized as an integrated first-class model that can be directly deployed to realize self-adaptation at runtime. ActivFORMS requires that the feedback loop model is: (i) \textit{verifiable}, i.e., the model can be used together with a \textit{model verifier} to check the correctness of the feedback loop behavior with respect to a set of correctness properties, and (ii) is \textit{executable}, i.e., the model specifies the behavior of the MAPE workflow such that it can be executed by a \textit{model execution engine} to monitor the managed system, reason about change, and adapt the system as needed, realizing self-adaptation. 

Designing a feedback loop for a problem at hand requires \textit{domain knowledge} (see Fig.~\ref{fig:activFORMS}). Domain knowledge refers to domain-specific data provided by stakeholders about the environment and the system itself that is relevant to adaptation. Examples are the behavior of users, the expected load of the system, initial values of the uncertainty parameters, and elements of the system that can be used to adapt the configuration of the system. The designer also requires a specification of the \textit{adaptation goals} that refer to the quality requirements that need to be realized by the feedback loop. ActivFORMS is not prescriptive in the types of adaptation goals supported, nor in the representation that is used to specify them. For a problem at hand, the designer needs to specify the adaptation goals in a format that allows the feedback loop model to reason about the goals at runtime. 

ActivFORMS relies on \textit{MAPE model templates} to design concrete feedback loops. These templates provide abstract designs of \textit{Knowledge} and \textit{MAPE models}. MAPE model templates consolidate design knowledge that is obtained from designing feedback loops for similar types of self-adaptive systems. The templates offer common elements of different MAPE models together with placeholders for application-specific elements of a feedback loop model. These placeholders need to be instantiated for the adaptation problem at hand. The Knowledge contains domain-specific models that are shared by the MAPE models, including models of the managed system, its environment, and quality models. For the managed system and the environment models the designer uses domain knowledge to identify the characteristics that are relevant for adaptation, including the current configuration of the system, the adaptation goals, adaptation options, relevant quality properties, and possibly other information. Domain-specific knowledge is also required to specify quality models, one for each adaptation goal. At a given point in time, when adaptation is required, the feedback loop needs to select an adaption option among the possible options. To select an adaptation option that satisfies the adaptation goals the system needs to know what the impact would be of the different options when applied on the managed system. The quality models allow finding this out. Quality models have two types of parameters: (1) settings of the managed system that determine the adaptation options, and (2) uncertainties of the managed system and its environment. By assigning values to these parameters, the feedback loop uses the quality models to determine what would be the expected quality values of the system if we choose any particular adaptation option. Based on the quality values for each adaptation option the feedback loop then chooses the best option based on adaptation goals.

The templates need to be supported by guidelines that describe how the templates can be instantiated for a given setting. Using MAPE model templates of an instance of ActivFORMS can significantly \mbox{reduce the effort to design feedback loop models and verify the correctness of their behavior.}

\subsubsection{Design Feedback Loop Model with ActivFORMS-ta}

ActivFORMS-ta supports the design of feedback loop models by a set of concrete MAPE model templates. The MAPE model templates are derived from extensive experience with modeling MAPE-based feedback loops for various applications, see e.g.,~\cite{Didac2013,Iftikhar2014,Shevtsov2015,Weyns2015c,7573167,Calinescu2017}. The common characteristics of these applications determine the types of systems targeted by the model templates, i.e.: (i) the managed system is available and instrumented with probes and effectors, and (ii) the dynamics faced by the system are such that the feedback loop has sufficient time to make adaptation decision (see also the general assumptions of ActivFORMS listed in Section~\ref{section:overview}). 
The model templates of ActivFORMS-ta are specified as a network of timed automata, and properties are specified in Timed Computational Tree Logic (TCTL). The Uppaal tool suite is used for the specification and verification of feedback loop models~\cite{Behrmann2004,David2015}. Template elements that apply to any self-adaptive system do not require instantiation, while other elements need to be instantiated  for the adaptation problem at hand (e.g., a function, guard, or property). We start with the design of the knowledge part. Then we zoom in on the MAPE models. We conclude with the rules and process to instantiate the model templates of ActivFORMS-ta.  

\paragraph{\textbf{Knowledge}} The MAPE model templates offer an abstract specification of the knowledge that consists \mbox{of the $\mathit{current}$ $\mathit{configuration}$, $\mathit{adaptation}$ $\mathit{goals}$, $\mathit{adaptation}$ $\mathit{options}$, a $\mathit{plan}$, and $\mathit{quality}$ $\mathit{models}$.} 

The current configuration represents the aspects of the managed system and its environment relevant to adaptation. The adaptation goals define the quality requirements that need to be realized by the feedback loop. ActivFORMS-ta supports modeling adaptation goals as boolean functions. An $\mathit{optimization}$ $\mathit{goal}$ tests whether a configuration outperforms a given configuration regarding a given property, while a $\mathit{satisfaction}$ $\mathit{goal}$ tests whether a configuration satisfies a given property. An adaptation option is determined by a particular setting of the managed system and is provided with a placeholder for the $\mathit{verification}$ $\mathit{results}$, i.e., the estimated values for the different qualities for that adaptation option produced by the verifier. A plan consists of a series of $\mathit{steps}$, each defined by a $\mathit{type}$, an  $\mathit{element}$ of the managed system that is subject to adaptation via the step, and a $\mathit{new}$ $\mathit{value}$ that needs to be applied to the element. Finally, a quality model is a domain-specific abstraction of the behavior of the managed system and its environment that captures the characteristics of a quality property. Quality models are specified as a parameterized stochastic timed automata. 
For a complete specification of the knowledge in the Uppaal language, we refer to the ActivFORMS website. 
\vspace{5pt}\\
\noindent
\textbf{Example 1.}  We illustrate the instantiantiation of the knowledge for DeltaIoT, see Listing~\ref{knowledgeDeltaIoT}. 
The current configuration of DeltaIoT is specified as the network of motes with their actual settings (i.e., the transmission power of the motes, distributions of packets to parents), the current values of quality properties (power loss and energy consumption), and uncertainties (current traffic load of motes and $\textit{SNR}$ of links). 
The packet loss goal is specified as a satisfaction goal and the energy goal as and optimization goal. The packet loss goal tests whether the packet loss of a configuration is not higher as a given threshold (here defined at 10\%). The energy consumption goal tests whether the energy consumption of one configuration is lower as that of another, allowing to find the configuration with the lowest energy consumption. An adaptation option is determined by particular settings of the network, i.e., the transmission power and distribution of packets of links. The verification results provide estimated values for packet loss and energy consumption for the adaptation option.
Two types of steps for plans are specified: change the power settings of a mote, e.g., $\textit{\{CHANGE\_POWER, mote7Id, link1Id, 5\}}$ says that the transmission power of mote~7 on link~1 is set to 5; and change the distribution of packets sent to parents, e.g., $\textit{\{CHANGE\_DISTR, mote7, link3, 60\}}$ says that mote~7 will send 60\,\% of its traffic via link~3 (i.e., to mote~3). 

\lstset{caption={Knowledge definition for DeltaIoT feedback loop.},label=knowledgeDeltaIoT}
\footnotesize
{\ttfamily
	\begin{lstlisting}
Configuration currentDeltaIoTConfiguration; //For details, see ActivFORMS website. 
	
//Adaptation Goals
int MAX_PACKET_LOSS = 10; //max packet loss 10%
bool satisfactionGoalPacketLoss(Configuration gConf, int MAX_PACKET_LOSS) {
  return gConf.qualities.packetLoss < MAX_PACKET_LOSS;
}
bool optimizationGoalEnergyConsumption(Configuration gConf, Configuration tConf) {
  return tConf.qualities.energyConsumption < gConf.qualities.energyConsumption;
}
	
//Adaptation Options
ManagedSystem deltaIoT_1 {...};  
Qualities verificResults_1 = {...};  ... 
AdaptationOption adaptationOptions[MAX_OPTIONS] = 
{{deltaIoT_1, verificResults_1}, {deltaIoT_2, verificResults_2}, ...};
	
//Plan with Step Types 
const CHANGE_POWER; 
const CHANGE_DISTRIBUTION;
//For the detailed definitions of motes and links, see Appendix B
Step step_1 = {CHANGE_POWER, mote7Id, link1Id, 5}; ...
Step step_4 = {CHANGE_DISTRIBUTION, mote7Id, link3Id, 60};  ...  
Plan plan  = {step_1, ... step_4, ... }
	
//Quality Models
//Models for packet-loss and energy consumption (networks of timed automata)
	
\end{lstlisting}
}
\normalsize

\begin{figure}[h!tb]
	\centering
	\includegraphics[width=0.8\textwidth]{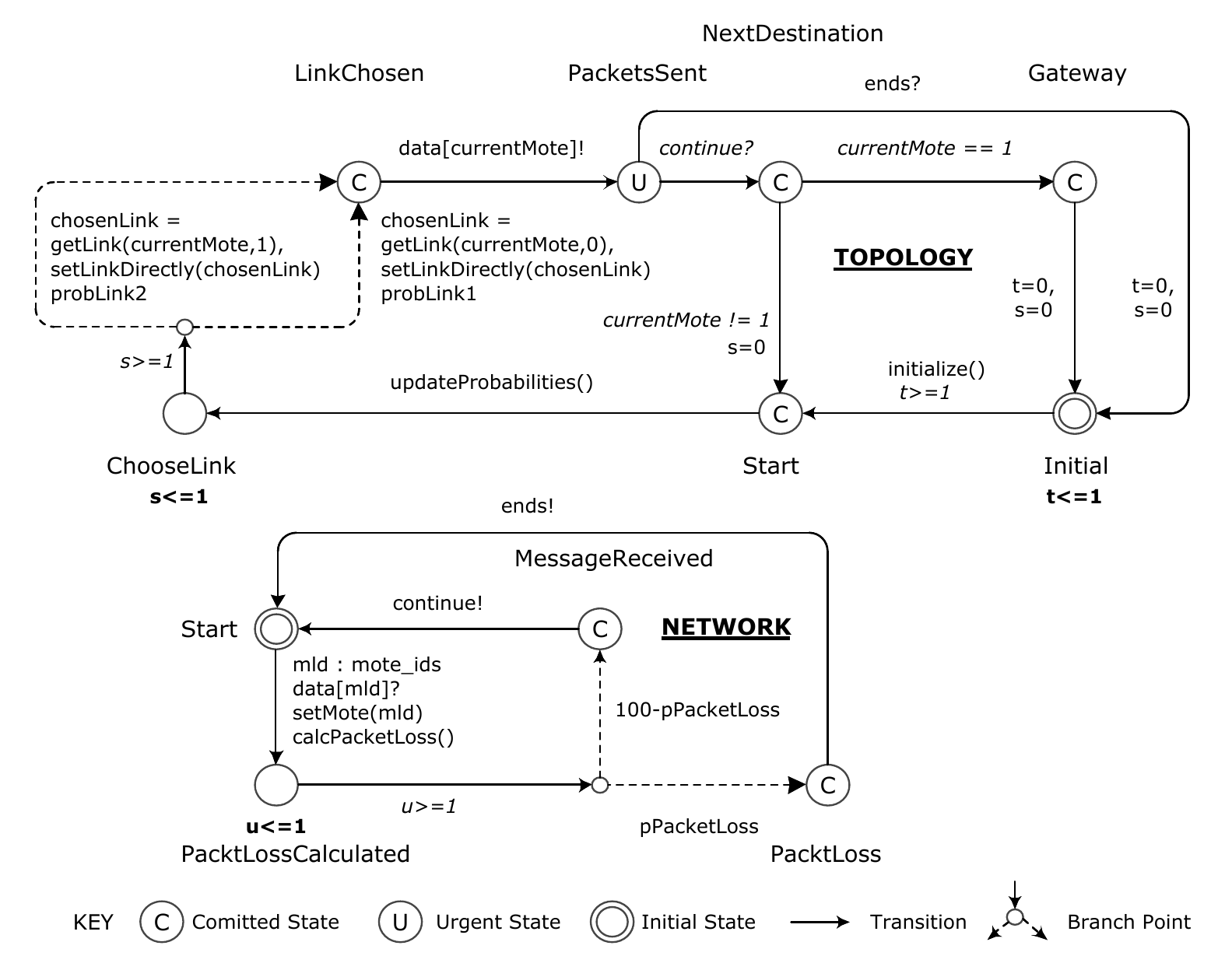}
	\caption{Runtime quality model for packet loss}
	\label{fig:runtime_quality_model_packet_loss}
\end{figure}

For the basic version of DeltaIoT, the designer needs to specify a quality model for packet loss and one for energy consumption. While these models have to be designed and tested in Stage I, they are only active at runtime to support the analysis of the adaptation options. Fig.~\ref{fig:runtime_quality_model_packet_loss} shows the quality model for packet loss shown that consists of two interacting automata: Topology and Network.

The behavior starts with the initialization of the model. The adaptation options are determined by the settings of the power and the distributions factors of the network. The power setting (0 to 15) for each link is set to the minimum value that is required to ensure that the SNR is at least zero, ensuring a low packet loss.\footnote{The SNR is a linear function of the power setting. We refer to the ActivFORMS website for a discussion in detail.} The distribution factors for links of motes with two parents are set from 0 to 100\% in steps of 20\% (0,100), (20,80) ... (100,0). These values are assigned to the variables $probLink1$ and $probLink2$ of the topology model. For motes with one parent that link is used for transmitting all packets.  
Furthermore, the values of the uncertainties, network interference (SNR) and traffic load, are set. These values are based on the recent observations and apply to all adaptation options. Note that some motes generate a steady traffic (e.g., periodic samples of the temperature, see Fig.~\ref{fig:DeltaIoT}), while other motes generate a fluctuating traffic (i.e., based on presence of humans).

After initialization, the \textit{Topology} automaton starts sending data along the path selected for verification, i.e. a sequence of links from one mote via other motes to the gateway (see also Fig.~\ref{fig:DeltaIoT}). 
The current link to send data is selected probabilistically based on the distribution factors ($probLink1$ and $probLink2$). The model then signals the \textit{Network} automaton. Next, the probability for packet loss is calculated, based on the recent value of the SNR. Depending on the result either the packet got lost or it was received. In the latter case, the network automaton returns to the start location, continuing with with the next hop of the communication along the path that is currently checked, until the gateway is reached. If a packet got lost, the communication along the path that is checked ends. As such, the quality model allows determining the packet loss of the adaptation options by performing simulations of the communication of packets through the network taking into account the current uncertainties until results with the required accuracy and confidence are obtained.

\paragraph{\textbf{MAPE Models}} 

\begin{figure}[h!tb]
	\centering
	\includegraphics[width=\textwidth]{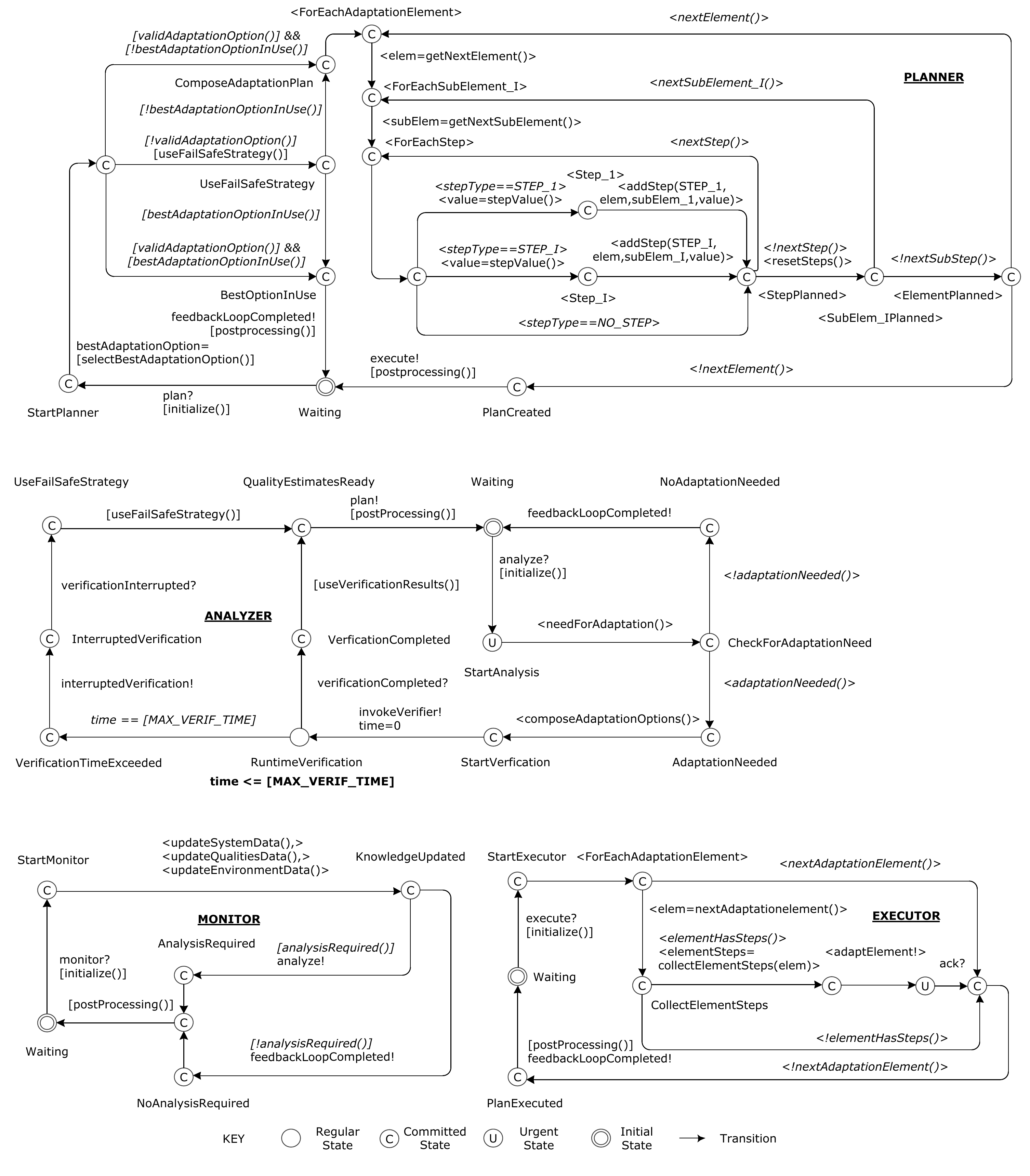}
	\caption{Reusable templates for specifying MAPE models}
	\label{MAPE_templates}
\end{figure}

When designing MAPE models for a problem at hand, the designer can use the abstract MAPE models of the MAPE model templates that are shown in Fig.~\ref{MAPE_templates}.\footnote{As a convention, elements in \textit{square brackets} are abstractly defined and need to be implemented (e.g., a function or a guard) without changing the name of the element. For elements in \textit{angle brackets} the same applies, but, these elements can be given domain-specific names. Domain-specific names in model templates support readability, but require a corresponding instantiation in the verification properties. Some domain-specific elements are marked as $name\_I$; these optional elements can be instantiated an arbitrary number of times.} The templates use event triggering, e.g., the monitor triggers the analyzer when analysis is required. ActivFORMS-ta also provide time triggered templates, where MAPE functions can be activated by an internal clock (the two versions of the MAPE model templates are available at the ActivFORMS website). 

\begin{figure}[h!tb]
	\centering
	\includegraphics[width=\textwidth]{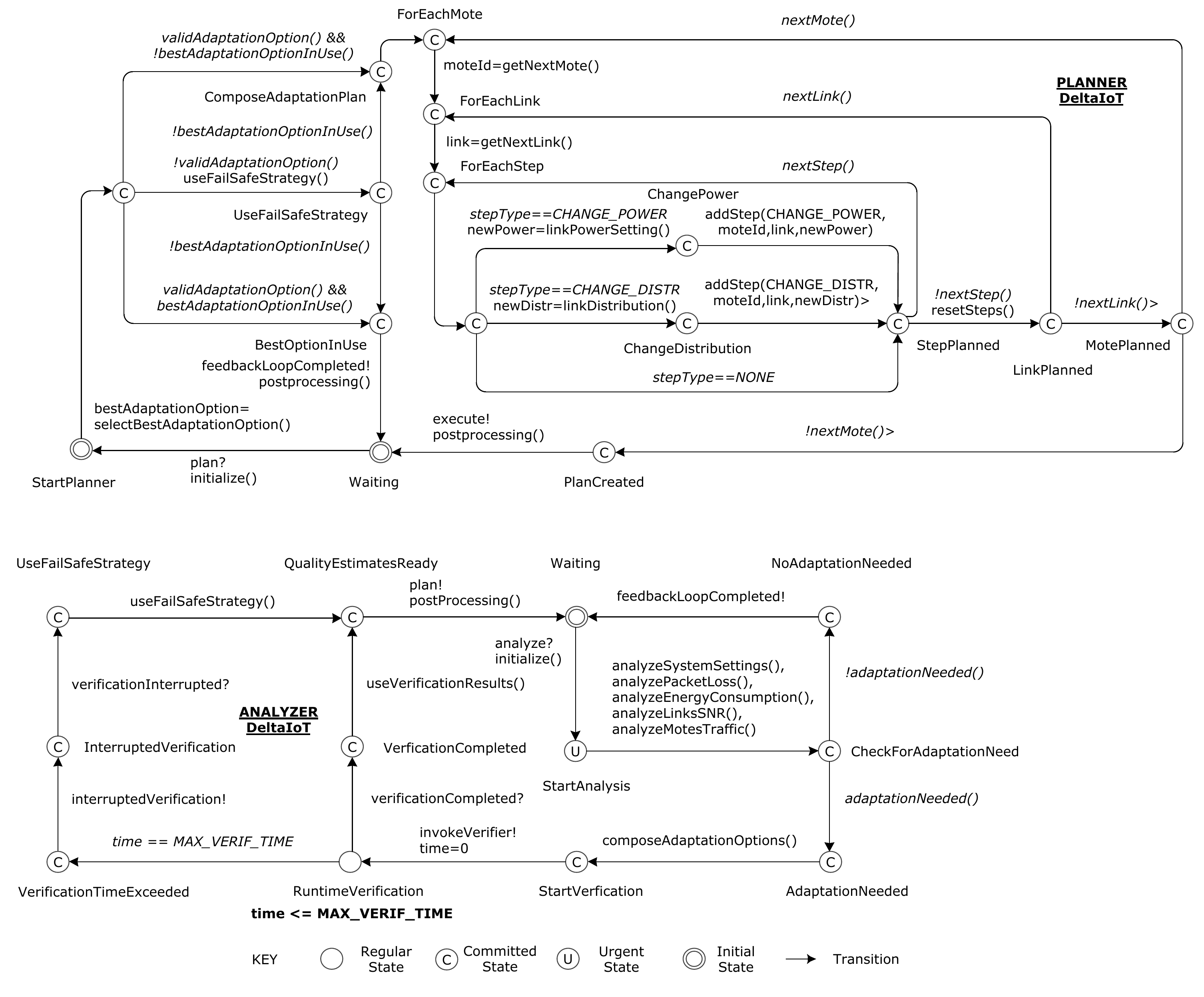}
	\caption{Two template instances for the MAPE models of DeltaIoT}
	\label{DeltaIoT_MAPE}\vspace{-5pt}
\end{figure}

When instantiating the MAPE models, the main tasks the designer needs to ensure are the following. When receiving new data, the \textit{Monitor} needs to update the knowledge (i.e., the parameters of uncertainties and quality properties of the system, see e.g.,\,\cite{7355393}), and check whether analysis is needed. If so, the \textit{Analyzer} needs to check whether adaptation is required, typically when adaptation goals are violated. If that is the case, the adaptation options need to be composed and verified (i.e., quality estimates are computed using the quality models) and subsequently the planner is triggered. In case the verification exceeds the maximum verification time, a failsafe adaptation strategy needs to be applied. The \textit{Planner} needs to select the best configuration by applying the adaptation goals to the adaptation options based on their quality estimates. The planner then composes a plan step by step. To that end, the planner identifies the elements and possibly  sub-elements that need to be adapted based on the difference between the current and new configuration. When all the steps of all elements are added to the plan the \textit{Executor} needs to be triggered. For each element, the executor collects all the plan steps associated with the element (and possibly its sub-elements) and triggers the effector to apply the adaptation actions to the managed system. This completes the MAPE workflow. 
\vspace{5pt}\\
\noindent
\textbf{Example 2.} Fig.~\ref{DeltaIoT_MAPE} shows instances of the templates for the analyzer and planner models of DeltaIoT.  The instantiations for the other MAPE models are available at the ActivFORMS website. 

To instantiate the \textit{Analyzer}, the designer needs to implement the abstract functions of the template model. Determining the need for adapting the IoT motes includes five functions. The function $\mathit{analyzeSystemSettings()}$ checks whether the network settings (power and distribution per link) are different from the expected settings (as applied in the last adaptation step). A difference indicates that the last adaptation steps were not effected as expected or the settings changed for another reason. The functions $\mathit{analyzePacketLoss()}$ and $\mathit{analyzeEnergyConsumption()}$ check whether the packet loss and energy consumption have increased significantly. Similarly, $\mathit{analyzeLinksSNR()}$ and $\mathit{analyzeMotesTraffic()}$ check the changes of $\textit{SNR}$ of the links and the traffic load generated by the motes. If at least one of the analysis functions returns true, $\mathit{adaptationNeeded()}$ returns true. 
If adaptation is needed, the adaptation options are composed as follows. The analyzer determines the power setting that is required per mote for each link. These settings are selected such that the $SNR_l \geq 0$  based on (1) link-specific functions $SNR_l$=$f(power_l)$.\footnote{The ActivFORMS website provides link-specific functions of the IoT network that we use in the evaluation.} 
The analyzer then determines all the possible combinations of packet distributions for all links (in a range of 0 to 100\% in steps of $20\%$) (motes with one parent send all packets to the parent). Each of these combinations determines an adaptation option. For a network with 15 motes as shown in Fig. \ref{fig:DeltaIoT}, this results in 216 adaptation options. If the network structure does not change, the number of adaptation options does not change. 

The designer needs to set $\textit{MAX\_VERIF\_TIME}$ for the DeltaIoT configuration at hand (e.g., for a deployment with a cycle time of $9.5$ minutes, the max verification time is set is to $8$ minutes). The function $\mathit{useVerificationResults()}$ is used to copy the estimated values for packet loss and energy consumption as determined by the verifier to the respective placeholders of all adaptation options. 
The following failsafe strategy can be used: if the partial verification results contain at least one adaptation option that satisfies the adaptation goals the best option is selected among these; if there is no such option, the settings of the reference approach are applied with maximum power settings for each mote and all motes send their  packets to all parents. 

To instantiate the \textit{Planner} model, the designer needs to instantiate the abstract elements of the model template and use the standard procedure of the template to realize the planning. To select the best adaptation option, the planner applies the adaptation goals to each of the adaptation options based on the quality estimates of the adaptation options. For the basic case with two goals, $\mathit{selectBestAdaptationOption()}$ first selects all adaptation options with an estimated packet loss below the threshold. From this subset, the option with minimum energy consumption will be picked for adaptation. If no such option is found, the fail safe strategy will be selected. If planning is required, the system determines the steps of the plan per mote and per parent link. The plan steps are $\mathit{ChangePower}$ that adapts the transmission power of a $\mathit{link}$ with a $\mathit{newPower}$ value, and $\mathit{ChangeDistribution}$ that adapts the percentage of packets sent over a link to a parent with a value $\mathit{newDistr}$.  

\subsection{Verify Feedback Loop Model} 

\subsubsection{Verify Feedback Loop Model with ActivFORMS}

Before deployment, the knowledge models need to be tested and the MAPE models need to be verified for correctness using stub models.

\paragraph{\textbf{Test Knowledge Models}}

The knowledge models that are used at runtime to predict the qualities of the adaptation options are domain-specific. Hence, dedicated testing is required. The purpose of testing is to provide the necessary evidence that the knowledge models make appropriate predictions. Setting up tests requires domain information about the adaptation options of the managed system, the uncertainties the system will be exposed to, and the qualities expected for the adaptation options under different conditions. Domain experts can derive such information from field tests or from another source. Based on this information, the designer can define a test strategy, including the tools to be used, the coverage of the tests, test setups with the necessary input, and the required accuracy of the results. Testing is usually done incrementally, where the models can be fine tuned based on the results until they are satisfactory, i.e., they make appropriate predictions.

\paragraph{\textbf{Design Stubs}}

Verifying the behavior of the MAPE models requires \textit{stubs} that represent abstractions of the behavior of the elements that the MAPE models interact with (see Fig.~\ref{fig:activFORMS}). These stubs are domain-specific and can be derived the from a specification or the implementation of the managed system and other elements the feedback loop model interacts with. The coverage of the verification results depends on the specification of the stubs, hence they should represent the behaviors of the corresponding external elements as required. To that end, the designer needs to ensure that the stubs generate the necessary input to verify the different behaviors of the MAPE models such that all the necessary paths of the models are exercised when verifying the respective properties. To ensure that the behavior of the stubs is compliant with the behavior of the external elements, the designer can use different techniques, e.g., model-based testing~\cite{Tretmans:2008} that checks the equivalence between the runtime behavior of software under test and the outcome generated by a model. It is the task of the designer to apply these general guidelines when designing the stub models for the problem at hand. 

\paragraph{\textbf{Verify MAPE Models}}

Besides generic models, the MAPE model templates of a concrete instance of ActivFORMS can also offer a set of generic properties that represent correctness requirements that should be satisfied by any feedback loop; an example is deadlock freeness. In addition, domain-specific correctness properties may be defined. Properties can be defined that check the correctness of a particular MAPE model, the interaction between MAPE models, and the overall behavior of the complete MAPE workflow. The properties need to be specified in a language that allows a verifier to check that the MAPE models behave correctly with respect to the properties. 
To verify the MAPE models, the designer needs to instantiate stub models with domain-specific data and connect the stubs to the models before verification starts. Stubs are required for the probes, effectors and the verifier, and possible other external elements the MAPE models are connected to. 

An important property of a feedback loop is ensuring that failsafe operating modes are always satisfied. To that end, a concrete instance of ActivFORMS needs to define properties for failsafe operation that the designer needs to instantiate for the problem at hand. Ensuring these properties guarantees that the adaptive system can switch to a fall-back or degraded operating mode when needed during operation. Note that instead of falling back to a failsafe strategy in case the goals cannot be achieved, the designer may add domain-specific logic to the analyzer to handle situations where some of the goals can be satisfied but not all of them.

\subsubsection{Verify Feedback Loop Model with ActivFORMS-ta}\label{subsubsec:verify_MAPE_models}

\paragraph{\textbf{Test Knowledge Models}}

Since the knowledge models are domain-specific, dedicated testing is needed that relies on domain information. Examples are expected values for uncertainties, representative configurations, expected output for given input, etc. This information is used to specify test automata that are used to test the models. In ActivFORMS-ta, the knowledge part is specified in the Uppaal language and quality models are specified as networks of stochastic timed automata. After connecting and instantiating the test automata, the knowledge can be tested using the Uppaal suite. 
\vspace{5pt}\\
\noindent
\textbf{Example 3.} We illustrate a test of the quality model for packet loss (see Fig.~\ref{fig:runtime_quality_model_packet_loss}). We used the Uppaal-SMC tool~\cite{David2015} to test the following query: 

\begin{quote}
	$Pr[<= 1] (<> Network.PacketLoss)$\vspace{-2pt}
\end{quote}

This query checks the probability of packet loss using a test automaton with the standard settings for the parameters of the network as provided by domain experts. The verification result is: 

\begin{quote}
	\textit{(175 runs) Pr($<>$ ...) in [0.0712232, 0.170976]
		with confidence 0.95.}\vspace{-2pt}
\end{quote}

\begin{figure}[h!tb]
	\centering
	\includegraphics[width=0.85\textwidth]{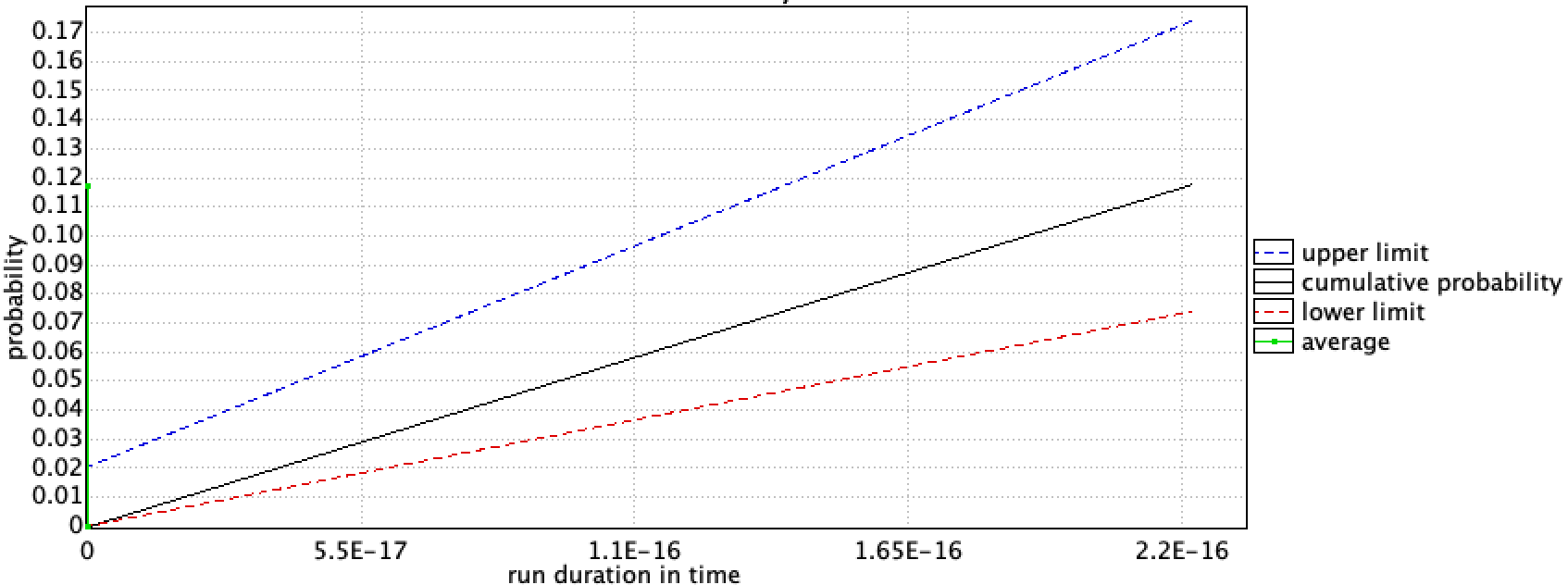}
	\caption{Cumulative probability of packet loss for a test setting of DeltaIoT}
	\label{data}\vspace{-5pt}
\end{figure}

Fig.~\ref{data} shows an excerpt of simulation results of the query for the cumulative probability of the packet loss with upper and lower boundaries. The average probability of packet loss of $0.12$ can then be compared with the input of the domain experts and if needed the models can be tuned and retested. 

By applying different settings of uncertainty parameters, the verification results will evaluate the correctness of the knowledge models for a broad range of conditions. 

\paragraph{\textbf{Design Stub Models}}

\begin{figure}[h!tb]
	\centering
	\includegraphics[width=\textwidth]{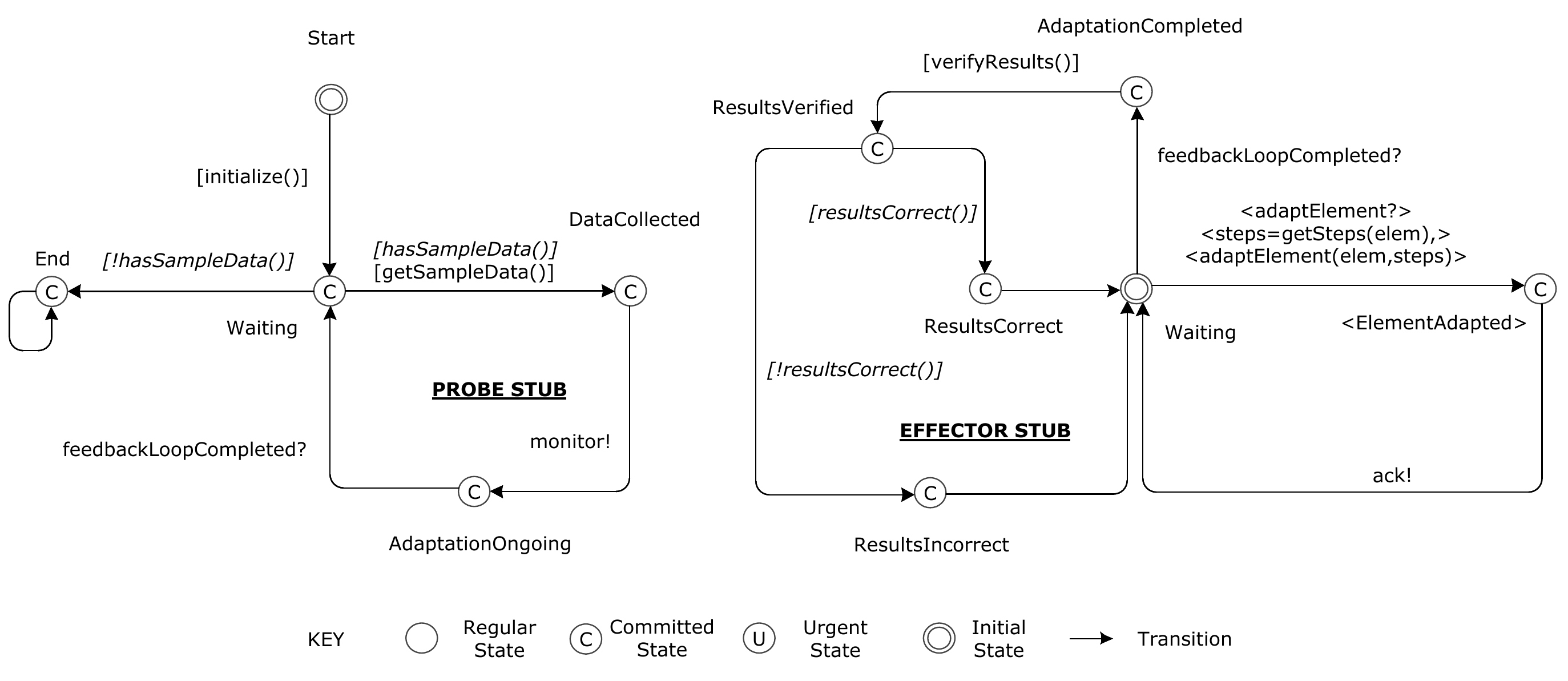}
	\caption{Templates for probe and effector stubs}
	\label{Stub_templates}
\end{figure}

Specifying stubs is a domain-specific effort. However, ActivFORMS-ta supports designers with a set of templates to devise the stubs for a problem at hand, specified as timed automata. Fig.~\ref{Stub_templates} shows the templates for the probe and effector stubs. 

For the \textit{Probe stub}, after initialization, the probe collects sample data from the system, the relevant qualities, and the environment. The sample data is typically specified as a sequence of configurations. The probe then triggers the monitor model of the feedback loop that starts an adaptation cycle. When the feedback loop cycle completes, the probe starts a new cycle as long as sample data is available. 

For the \textit{Effector stub}, when an adaptation action is invoked by the executor model, the effector determines the element of the managed system that needs to be adapted and the steps that need be be applied. Once the feedback loop workflow is completed, the effector receives a notification from the MAPE models; it can then
check whether the configuration is correctly adapted (\textit{ResultsCorrect}) or not (\textit{ResultsIncorrect}).
For more info about stubs for DeltaIoT, we refer to the ActivFORMS website.

\paragraph{\textbf{Verify MAPE Models}}\label{section:properties}

ActivFORMS-ta offers a set of generic properties that the MAPE models specified with the model templates should comply to. These properties refer to functional requirements for the feedback loop and are pivotal in ensuring correct behavior of the feedback loop. The properties are specified in TCTL, using the Uppaal modeling language~\cite{tutorial04} (for the grammar, see the ActivFORMS website). As explained in Section \ref{section:preliminaries}, TCTL expressions allow verifying properties such as safety, liveness, etc. The Uppaal tool is used to verify the MAPE models. ActivFORMS-ta offers the following set of basic properties: 

\footnotesize
\begin{verbatim}
  P1. Probe.DataCollected --> Monitor.KnowledgeUpdated
  P2. Monitor.AnalysisRequired --> Analyzer.CheckForAdaptationDone
  P3. Analyzer.AdaptationNeeded --> Verifier.VerificationDone 
  P4. Analyzer.QualityEstimatesReady --> 
      Planner.ComposeAdaptationPlan || Planner.BestOptionInUse
  P5. Analyzer.VerificationTimeExceeded -->    
      Analyzer.UseFailSafeStrategy   
  P6. Planner.PlanCreated --> Executor.PlanExecuted
  P7. Executor.PlanExecuted --> Effector.AdaptationCompleted
  P8. Planner.<ElementPlanned> && <Planner.elemId == e> && 
      Planner.<stepsContains(e, STEP_I, val)> -->
        Executor.<AdaptElement> && <Executor.elemId == e> &&
        Executor.<stepsAppliedContains(e, STEP_I, val)>
  P9. Executor.<AdaptElement> && <Executor.elemId == e> &&
      Executor.<stepsAppliedContains(e, STEP_I, val) --> 
        Effector.<ElementAdapted> && <Effector.elemId == e> &&
        Effector.<stepsEnactedContains(e, STEP_I, val)>      
  P10. A[] !Effector.ResultsIncorrect
  P11. E<> <Model.Location> 
  P12. A[] no deadlock
\end{verbatim}
\normalsize

Properties Pr1 to Pr7 have obvious semantics. While these properties seems trivial when observing the MAPE models (see Fig.~\ref{MAPE_templates} and Fig.~\ref{DeltaIoT_MAPE}), it is important to note that verifying these properties allows checking the correct instantiation of the underlying domain-specific logic of the model templates (functions, guards, etc.). 
Properties $P8$ and $P9$ state that the steps to adapt an element generated by the planner are eventually applied by the executor and then enacted by the effector. 
Property $P10$ states that location $\mathit{ResultsIncorrect}$ of the $\mathit{Effector}$ model is never reached (see above). This property allows checking that the MAPE models perform the adaptation of a feedback loop cycle correctly. Property $P11$ on the other hand states that there exists a path to a given $\mathit{Location}$ of a given $\mathit{Model}$. Both location and model are abstractly defined and can be instantiated for the domain at hand. Property $P12$, which is supported by Uppaal, allows verifying whether the system is deadlock free. Elements in angle brackets need to be instantiated according to the domain-specific MAPE models. 
\vspace{5pt}\\
\noindent
\textbf{Example 4.} To verify the MAPE models for DeltaIoT, the designer needs to connect the MAPE models with the stub models of the probe and effector, and the verifier (for the analyzer and planner models, see Fig.~\ref{DeltaIoT_MAPE}; for the other models, we refer to the ActivFORMS website). The integrated model can then be verified. Properties $P1$ to $P7$ and $P10$ and $P12$ can be directly applied to the MAPE models. Properties $P8$, $P9$ and $P11$ on the other hand need to be instantiated by the designer. We illustrate this instantiation with examples for $P8$ and $P11$: 

\footnotesize
\begin{verbatim}
  //Generic property 
  P8. Planner.<ElementPlanned> && <Planner.elemId == e> && 
      Planner.<stepsContains(e, STEP_I, val)> -->
        Executor.<AdaptElement> && <Executor.elemId == e> &&
        Executor.<stepsAppliedContains(e, STEP_I, val)>
  //Concrete instance for DeltaIoT  
  P8c. Planner.MotePlanned && Planner.moteId == mote2Id && 
       Planner.stepsContains(mote2Id, link1Id, CHANGE_POW, 5) -->
         Executor.AdaptMote && Executor.moteId == mote2Id && 
         Executor.stepsAppliedContains(mote2Id, link1Id, CHANGE_POW, 5)
  //Generic property
  P11. E<> <Model.Location>
  //Concrete instance for DeltaIoT (selected property)
  P11c. E<> Planner.UseFailSafeStrategy 
\end{verbatim}
\normalsize

Property $P8c$ checks that if the planner has planned the steps to adapt the settings of mote 2 (with $\mathit{mote2Id}$) and these steps include a step to change the power setting ($\mathit{CHANGE\_POW}$) of link 1 (with $\mathit{link1Id}$) to a setting $5$, this step will eventually be applied by the executor. $Pr11c$ checks whether a path exists to location $\mathit{UseFailSafeStrategy}$ of the $\mathit{Planner}$ model. Instantiating property $P11c$ allows checking whether the input used for verification is complete, i.e., all paths of the models are traversed. 

When the domain-specific properties are specified, they can be verified. 
We present the results in Section~\ref{section:evaluation}. For more details about property verification of DeltaIoT, see the ActivFORMS website. 

\subsection{Rules for Instantiating the MAPE Model Templates in ActivFORMS-ta}\label{subsubsec:rules_usage_templates} 

To instantiate the MAPE model templates, ActivFORMS-ta defines a set of rules that need to be respected when instantiating the templates for a concrete adaptation problem. These rules cover the obligations and constraints for instantiating the template models. For the templates shown in Fig.\,\ref{MAPE_templates} the rules are defined as follows:

\begin{enumerate}
	\item Abstractly defined elements of model templates marked with square brackets need to be implemented for the problem domain at hand; the names of these elements cannot be changed. 
	\item Abstractly defined elements of model templates marked with triangle brackets need to be implemented for the problem at hand (multiple instances are possible); these elements can be given domain-specific names.
	\item The names of abstractly defined elements of property templates marked with triangle brackets need to correspond with the domain-specific names used in the models. 
	\item Elements of model templates that are marked as $\mathit{name\_I}$ represent a facultative model construct; these elements can be instantiated as many times as needed for the domain at hand.  
	\item The names of the elements of property templates that are marked as $\mathit{name\_I}$ need to correspond with the domain-specific names used in the models.  
\end{enumerate}  

Additionally, designers can refine transitions of particular model templates or extend models. To guarantee the correct behavior of these extensions the designer needs to specify and verify domain-specific properties. The designer can also remove parts of the model templates; the related template properties will then not apply. 

\subsection{Summary of guarantees offered by the ActivFORMS approach in Stage I} 
\vspace{4pt}
\begin{mdframed}
	\textit{Guarantees}: by formally specifying and verifying MAPE models, possibly using MAPE model templates, stage I guarantees the correct behavior of the MAPE models with respect to a set of correctness properties;\vspace{2pt}\\
	\textit{Scope}: the guarantees for the properties are confined to the behavior space of the MAPE models that is
	defined by the behaviors that are exercised by the stub models during verification of the different properties.
\end{mdframed}

\section{Stage II  ActivFORMS: Deploy and Enact Feedback Loop}\label{section:stage_II}

In Stage II of the ActivFORMS approach, the verified feedback loop model is deployed and enacted using a \textit{model execution engine}, see Fig.~\ref{fig:activFORMS}. We start with deployment, then we present enactment. 

\subsection{Deploy Feedback Loop Model}

\subsubsection{Deploy Feedback Loop Model with ActivFORMS}

One of the distinct features of ActivFORMS is direct deployment and execution of the verified feedback loop model to realize adaptation of the managed system using a \textit{model execution engine}. If this engine executes the feedback loop model correctly, i.e., according to the semantics of the modeling language, it ensures that the guarantees for the correct behavior of the feedback loop model obtained in the first stage are preserved. Direct model execution avoids manual model to code translation, which can be an error-prone activity, and it paves the way to flexible updates of the running feedback loop model. 

Guaranteeing that the model execution engine executes the feedback loop model correctly can be a labor intensive effort. Yet, this effort needs to be done only once. A concrete instance of ActivFORMS may use an off-the-shelf model execution engine that may come with guarantees or it may offer a dedicated execution engine for which guarantees need to be provided. Depending on the needs, different techniques can be used to provide such guarantees, ranging from testing to formal proof. 

Preparing the feedback loop model for execution (see Fig.~\ref{fig:activFORMS}) involves three steps. First, the developer needs to deploy the \textit{model execution engine} together with the feedback loop model. Deployment includes the instantiation, configuration, and installation of the software. Depending on the model execution engine that is used, this may require manual intervention or can be automated. The model execution engine typically translates the feedback loop model to an internal format that is then used for execution. Second, the feedback loop model needs to be connected to the managed system, which is realized through \textit{probes} and \textit{effectors}. Recall that ActivFORMS assumes that the managed system is available and instrumented with probes and effectors. ActivFORMS does not prescribe how the connection between the feedback loop model and the managed system is realized. A concrete instance of ActivFORMS may offer dedicated mechanisms to directly link the monitor model with probes and the executor model with effectors (e.g., a set of abstract classes that need to be instantiated for the domain at hand), or the designer needs to provide these links through a dedicated implementation. Third, a \textit{statistical model verifier} needs to be deployed and connected with the feedback loop model allowing the analyzer to estimate the qualities for the adaptation options to make decisions about how to adapt the system from its current configuration when needed. ActivFORMS does not prescribe a specific model verifier and how it is connected with the feedback loop. Similar to linking the probes and effectors, a concrete instance of ActivFORMS may offer dedicated mechanisms to link the analyzer model and a model verifier, or the developer needs to provide this link through an implementation. Optionally, additional external elements may need to be connected with the feedback loop model; e.g., a plug-in module to support planning.

Regardless of the type of mechanisms that are used, ensuring correct communication between the feedback loop model and the external elements is crucial. When the designer develops specific classes to realize the connections, such guarantees can be provided through extensive testing.

\subsubsection{Deploy Feedback Loop Model with ActivFORMS-ta}\label{sub:instantiateVM}

ActivFORMS-ta offers a trusted virtual machine to execute the feedback loop model specified as a network of timed automata; the trustworthiness relies on extensive testing, for the test report see the ActivFORMS website.

The deployed feedback loop model consists of the MAPE models together with the Knowledge models. The knowledge models include the quality models that are used to estimate the quality properties of the adaptation options and the adaptation goals that are used to select configurations for adaptation. When a feedback loop is loaded, the virtual machine transforms the models with their locations and edges to an internal graph representation. The labels on the edges and states, e.g., guards, invariants, etc. are converted to task graphs. A task graph consists of a list of tasks that need to be executed when activated, such as updating a variable, evaluating an expression, etc. Once the model is converted, the virtual machine initializes all the signals and assigns a unique identifier to each signal. The model is then prepared for execution. For details about the internals of the virtual machine of ActivFORMS-ta, we refer the interested reader to \cite{Iftikhar2016}. 

The feedback loop model can be connected with external elements through channels. ActivFORMS-ta provides a set of template classes to connect probes, effectors, and a statistical model checker with a feedback loop model. These template classes support engineers with implementing the connections for a problem at hand. To ensure that the communication between the external elements and the MAPE models is implemented correctly, the designer needs to test the instantiated classes.  

Realizing a connection to transfer data from the external element to the model boils down to: (1) connect the model with the external element via the relevant channels, (2) implement the logic to receive data from the element, (3) translate the received data to a format that the model understands, (4) send the data to the model. Realizing a connection to transfer data from the model to the external element consists of: (1) connect the model with the external element via the relevant channels, (2) implement the logic to receive data from the model, (3) translate the received data to a format that the element understands, (4) send the data to external element.

In addition to the template classes, ActivFORMS-ta offers a generic plug-in mechanism to attach external elements with the virtual machine. A concrete plug-in is the live update manager that enables runtime updates of a feedback loop model. We elaborate on this plug-in in Stage IV. 
\vspace{5pt}
\\
\noindent
\textbf{Example 5.}  We illustrate how the executor model of the DeltaIoT feedback loop is connected with the effector of the network. Listing~\ref{effector-connector} shows how the template class is used to realize the connection. 

The connector gets the identifiers of the channels to connect the model with the effector. The $\mathit{receive}$ method accepts adaptation actions and effect them on the managed system through the effector. The $\mathit{ack}$ signal acknowledges the actions to the executor model. For the connection of the feedback loop model with the probe and the model checker, we refer to the ActivFORMS website. 

\lstset{caption={Connecting the executor model with the DeltaIoT effector.},label=effector-connector}
\footnotesize
{\ttfamily
	\begin{lstlisting}
   public EffectorConnector(ActivFORMSEngine engine, Effector effector) {
    // Get channel identifiers from engine
    adaptMote = engine.getChannel("adaptMote");
    ack = engine.getChannel("ack");
    // Connect executor model with effector via adaptMote channel 
    engine.register(adaptMote, "mote", "linkSettings");
   }
   @Override
   public synchronized void receive(int channelId, HashMap data) {
    if (channelID == adaptMote){
    // effect mote settings through effector
    }
    //Acknowledge actions via ack channel
    engine.send(ack);
   }
	\end{lstlisting}
}
\normalsize

\subsection{Enact the Model Execution Engine}

When the model execution engine and the feedback loop model are deployed and the connections are established (with the probe, effector, and verifier), the model execution can be started. Depending on the used model execution engine, some configuration may be required before the execution can start. 

Enacting the virtual machine of ActivFORMS-ta is straightforward. Once all external connections are established and the models are initialized, the last task is to define the real time that corresponds to one logical time unit in the model. The virtual machine can then be started, enacting self-adaptation.
\vspace{5pt}\\
\noindent
\textbf{Example 6.} 
Listing~\ref{start-VM} shows the steps to start the virtual machine for DeltaIoT. 

\lstset{caption={Start the virtual machine for the DeltaIoT network.},label=start-VM}
\footnotesize
{\ttfamily
	\begin{lstlisting}
   //Load feedback loop model
   ActivFORMSEngine engine = new ActivFORMSEngine("/models/DeltaIoT-MAPE.xml");
   // Set model time unit to real time unit in milliseconds
   engine.setRealTimeUnit(1000);
   // Initialize connections
   ...
   // Start the virtual machine
   engine.start();
   \end{lstlisting}
}
\normalsize
The virtual machine starts with loading the feedback loop model of DeltaIoT. Then, the real time that corresponds with one time tick on the model is set to $\mathit{1000}$\,ms. Finally, when the external connections with the probe, effector, and the verifier are set, the engine is started. 

\subsection{Summary of guarantees offered by the ActivFORMS approach in Stage II} 

\begin{mdframed}	
	\textit{Guarantees}: by deploying and enacting the verified MAPE models using a model execution engine, stage II ensures that the guarantees for the behavior of the MAPE models obtained in stage I are preserved;\vspace{2pt}\\
	\textit{Scope}: the guarantees hold under the assumption that the model execution engine executes the MAPE models correctly, and the feedback loop model communicates correctly with the external elements.
\end{mdframed}

\section{Stage III ActivFORMS: Runtime Analysis and Decision-Making}\label{runtime}\label{section:stage_III}

Stages I and II of ActivFORMS focus on the design and deployment of a feedback loop model. The third stage is a runtime stage where the verified feedback loop model executed by the execution engine monitors the managed system and its environment and adapts the managed system to realize the adaptation goals (see Fig.~\ref{fig:activFORMS}).

Stage III complements the guarantees of stages I and II with evidence that the self-adaptive system selects adaptation options that guide the managed system to realize the adaptation goals. A distinct contribution of ActivFORMS is that it aims to perform this decision-making process in an efficient way, i.e., with limited resources and within limited adaptation time. To that end, ActivFORMS relies on statistical verification at runtime. Statistical verification allows the feedback loop system to select adaptation options that comply to the adaptation goals with a required accuracy and level of confidence. We start the explanation with a high-level overview of the runtime architecture of ActivFORMS that shows the composition of the runtime elements. Then, we zoom in on runtime analysis and decision-making using statistical verification at runtime.

\subsection{Runtime Architecture of the ActivFORMS Approach} 

\begin{figure}[h!tb]
	\centering
	\includegraphics[width=0.65\textwidth]{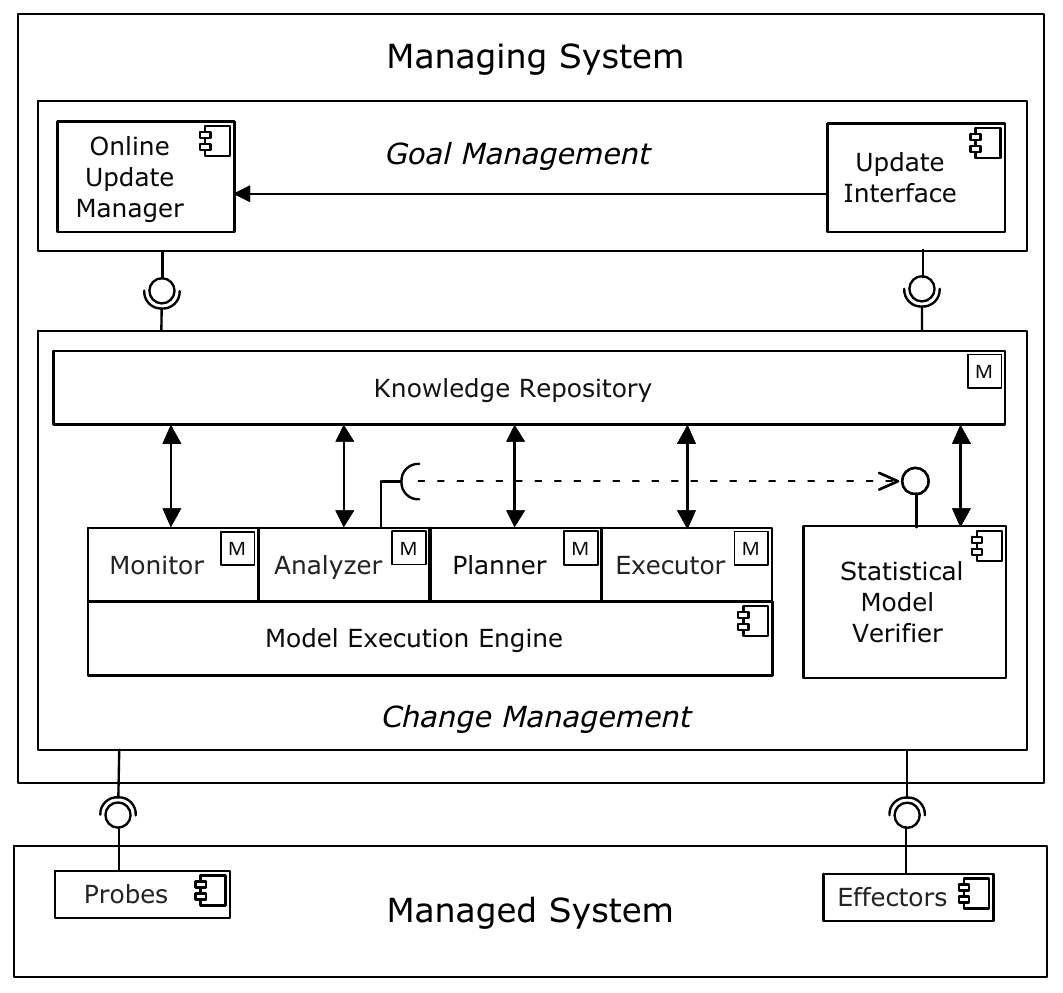}
	\caption{ActivFORMS runtime architecture (runtime models are marked with $\mathit{M}$).}
	\label{fig:activFORMS-RA}
\end{figure}

Fig. \ref{fig:activFORMS-RA} shows an overview of the ActivFORMS runtime architecture that aligns with the reference model for self-adaptive systems of~\cite{Kramer2007}. The \textit{Managed System} is the software that is subject of adaptation. In ActivFORMS, the managed system, instrumented with probes and effectors is given. At a given point in time the managed system has a configuration that is determined by the arrangement and settings of the running elements of the system. Adapting the managed system boils down to changing the configuration. The adaptation options define the set of  configurations that can be reached from the current configuration by adapting the managed system. This set is determined by combining the possible settings of different elements of the managed system that can be adapted. 

The \textit{Managing System} consists of two sub-layers. \textit{Change Management} comprises the verified \textit{MAPE models} that are executed by a model execution engine. The MAPE models share knowledge that is stored in the \textit{Knowledge Repository}, incl. a model of the managed system, quality models, and adaptation goals. The \textit{Monitor} updates the knowledge using data collected by the \textit{Probes}. The \textit{Analyzer} is supported by a \textit{Statistical Model Verifier} that runs simulations on the quality models to estimate the relevant qualities for the adaptation options. The \textit{Planner} uses the adaptation goals to select the adaptation option that complies best with the goals and create a plan to adapt the managed system accordingly. This plan is then executed by the \textit{Executor} via the \textit{Effectors}.

\textit{Goal Management} comprises an \textit{Online Update Manager} and an \textit{Update Interface}. The online update manager enables operators to update the feedback loop model during execution. Such updates are loaded by the update manager through the update interface. Changing the models needs to be done safely, i.e., in quiescent states~\cite{Kramer:1990}. We discuss goal management in detail in Stage IV.

We discuss now key activities of Stage III: analysis of  adaptation options and decision-making. 

\subsection{Analysis of the Adaptation Options}\label{subsec:analysis}

\subsubsection{Analysis of the Adaptation Options with ActivFORMS}\label{subsec:analysis}

The purpose of analysis is to provide estimates of the different quality properties for the adaptation options. In ActivFORMS, analysis is performed on first-class quality models. Different from existing approaches that in essence rely on exhaustive verification techniques, ActivFORMS applies statistical model verification. Compared to exhaustive verification, statistical model verification is more efficient in the time and resources required to perform the analysis. As explained in Section~\ref{section:statistical_model_checking}, the time required for statistical model verification is determined by the accuracy $\epsilon$ and confidence $\alpha$ of the verification queries (the time is polynomial in 1/$\epsilon$ and log 1/$\alpha$ \cite{Herault2004}). Compared to exhaustive verification, statistical model verification has the advantage that the accuracy and confidence parameters of verification queries can be set by the designer (and possibly be tuned during operation, manually or automatically). More accurate results with higher confidence result in better quality estimates, but increase the verification time. Hence, an important design decision is to balance accuracy and confidence of the analysis results with the time window that is available for verification at runtime (see also Section \ref{section:statistical_model_checking}).

Analysis consists of four steps: (1) composing the adaptation options by assigning values to the parameters that represent elements of the managed system that can be adapted, and providing for each adaptation option placeholders to store the estimates of the quality properties, (2) assigning values to the uncertainties stored in the knowledge repository by the monitor (typically as parameters in runtime models), (3) invoking the model verifier using the different quality models and the verification queries for each adaptation option (see also  Section~\ref{section:statistical_model_checking}), (4) updating the placeholders for the quality estimates of the adaptation options using the verification results. If no valid adaptation option can be found, e.g., when the verification time exceeds the available time window, the a failsafe configuration should be selected to adapt the managed system. This completes analysis.

\subsubsection{Analysis of the Adaptation Options with ActivFORMS-ta}\label{subsubsection_analysis_options}

To ensure that a self-adaptive system achieves its adaptation goals, ActivFORMS-ta uses statistical model checking, relying on Uppaal-SMC~\cite{David2015}. 
The analysis of the adaptation options (composed by the Analyzer, see also Stage I) relies on the quality models that are specified as networks of stochastic timed automata, one for each  quality that is subject of adaptation. The uncertainties in these models are represented as parameters that are initially assigned values based on input from domain experts and updated during operation based on observations of the monitor. 
Analysis provides estimates for the different quality properties of the adaptation options using statistical verification. To that end,  ActivFORMS-ta uses two types of verification queries: probability estimation ($p = Pr[bound](\varphi)$) and simulation ($simulate~N [\leq bound]\{E1,...,Ek\}$), see Section \ref{section:preliminaries}. For probability estimation, Uppaal automatically determines the number of simulation runs that are needed for the required accuracy and confidence. This is not possible for simulation queries. In this research we use the relative standard error of the mean (RSEM) as a measure to determine the accuracy of the simulation queries. The standard error of the mean (SEM) quantifies how precisely a simulation result represents the true mean of the population expressed in units of the data, taking into account the standard deviation and sample size. RSEM is the SEM divided by the sample mean and is expressed as a percentage. E.g., a RSEM of $1\%$ represents an accuracy with a SEM of $\pm 0.1$ for a mean value of $10$. Better estimates require smaller RSEM values and thus more simulation runs. RSEM provides a simple, but precise measure for estimating quality properties (and when the probability distribution is known, it allows calculating an exact confidence interval); however, ActivFORMS does not exclude using other measures.

We determine the number of simulations required for a given accuracy using off-line experiments. Concretely, we run simulations on a relevant set of samples for the domain at hand (e.g., a randomly selected set of 20\% of system configurations with randomly assigned values for uncertainty parameters in a given range) and empirically determine the number of runs that are required to obtain the required accuracy. If necessary, additional experiments can be run as a background process to deal with significant changes of models. In our current research, we rely on off-line experiments only.\footnote{The ActivFORMS website provides reports of the empirical experiments for DeltaIoT used in the evaluation section.} 
\vspace{5pt}\\
\noindent
\textbf{Example 7.} We illustrate how the designer realized the analysis of adaptation options for packet loss in DeltaIoT (standard setup). We presented the quality model for packet loss that consists of two interacting automata: Topology and Network in Fig.~\ref{fig:runtime_quality_model_packet_loss}. The feedback loop uses this model during operation to estimate the packet loss for the different adaptation options, using the following query:

\lstset{caption={Verification query for packet loss model},label=q-pl}
\footnotesize
{\ttfamily
	\begin{lstlisting}
			Pr [<=1](<>Network.PacketLoss)
	\end{lstlisting}
}
\normalsize

This query will perform a series of simulations and return an approximation interval $[p-\epsilon, p+\epsilon]$ with a confidence $1-\alpha$, where $p$ is the true packet loss and $\epsilon$ and $\alpha$ are set by the designer.

For the analysis of energy consumption and other models, we refer to the ActivFORMS website.

\subsection{Decision-Making}

\subsubsection{Decision-Making with ActivFORMS}

The goal of decision-making is to select the best option for adaptation. In ActivFORMS, the decision-making mechanism is realized by the planner that applies the adaptation goals to select the best adaptation option among the set of available options based on the verification results per adaptation option. ActivFORMS supports different types of adaptation goals, such as rules and utility functions, see e.g.,~\cite{2018Trollmann}. If no suitable adaptation option is available, e.g. none of the adaptation options satisfies a minimum number of adaptation goals, a predefined failsafe configuration is used to adapt the managed system. Once an option is selected, the planner creates a plan to adapt the managed system. 

\subsubsection{Decision-Making with ActivFORMS-ta}

In ActivFORMS-ta, decision-making is realized by the planner model, see Fig.~\ref{MAPE_templates} for the template of the planner model.  
ActivFORMS-ta implements two predefined types of adaptation goals: an optimization goal (returns the most optimal configuration of two given configurations for a given property), and a 
satisfaction goal (tests whether a given configuration satisfies a given property), see Listing~\ref{knowledgeDeltaIoT}. The designer can set the order in which the planner will apply the goals; the choice for this ordering depends on the adaptation problem at hand. Once the best adaptation option is selected, the planner composes a plan comprising a set of steps, one for each adaptation action that is required to adapt the managed system from its current configuration to the configuration of the selected adaptation option; see Fig.~\ref{MAPE_templates}. 
\vspace{5pt}\\
\noindent
\textbf{Example 8.}  We illustrate decision-making in DeltaIoT for a setting with 15 motes and two adaptation goals. Packet loss is defined as a satisfaction goal (packet loss $<10\%$) and energy consumption as a optimization goal (minimize energy consumption) For the definitions of the functions of the goals, we refer to Listing~\ref{knowledgeDeltaIoT}. First packet loss is applied, then energy consumption.   

\begin{figure}[h!tb]
	\centering
	\includegraphics[width=0.8\textwidth]{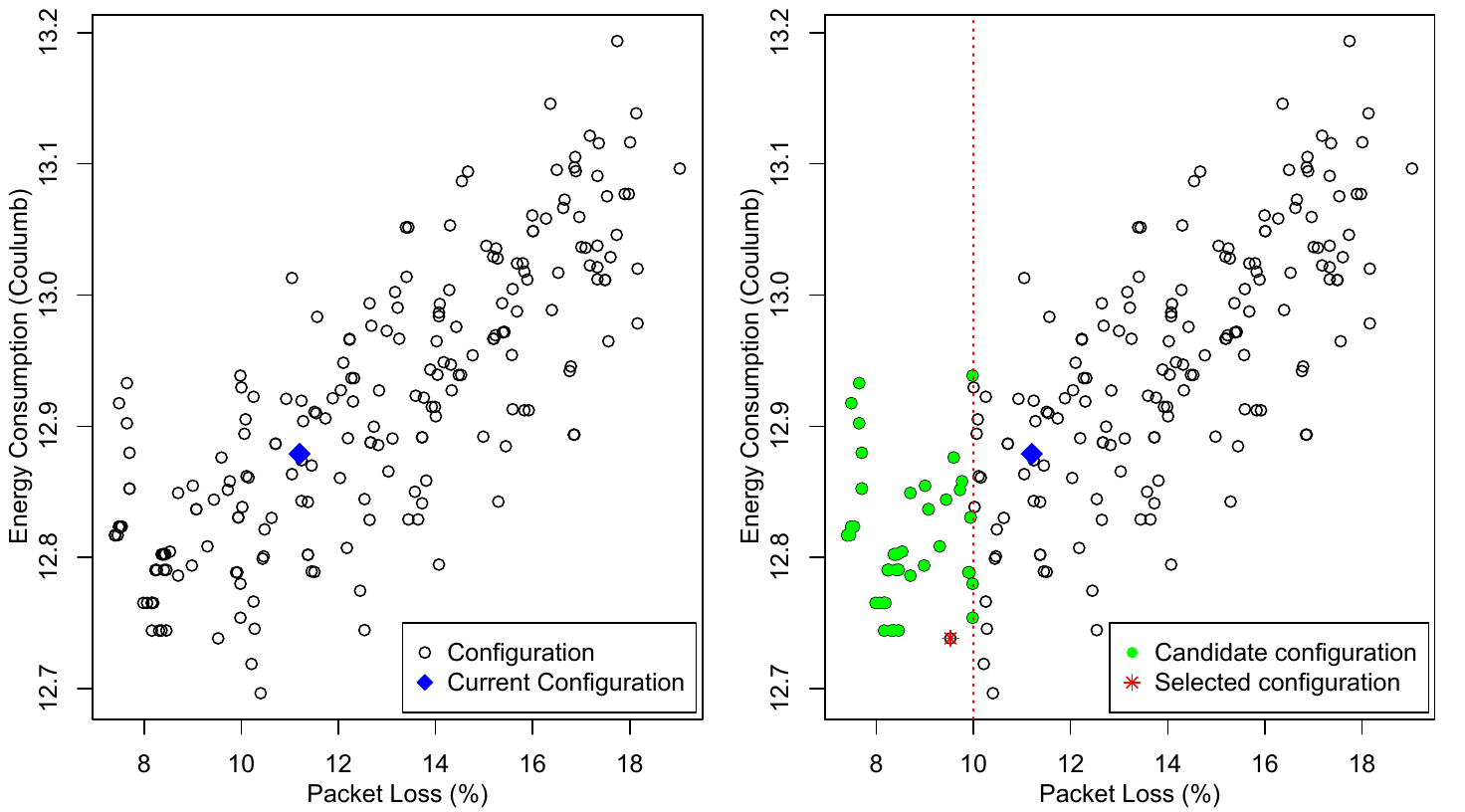}
	\caption{Decision-making at a particular point in time with two adaptation goals}
	\label{fig:decision_making}
\end{figure}

Fig.~\ref{fig:decision_making} shows an overview of the adaptation options at a particular point in time. Each dot on the left graph represents an adaptation option with its estimated average values of the two quality properties. The diamond dot (marked in blue) represents the configuration of the managed system in use at the time the analyse is performed. The shaded dots (marked in green) on the right graph represent adaptation options that comply with the adaptation goal for packet loss. Finally, the dot marked with a star (in red) on this graph represents the best adaptation option, i.e., the option with minimum energy consumption. This option is selected for adaptation and a plan is composed by the planner that adapts the current configuration to this new configuration. The selected adaptation option at this particular point in time is expected to reduce packet loss to $9.5\%$ and energy consumption to $12.75$~C compared to respectively $11.3\%$ and $12.88$~C of the current configuration.\footnote{Energy consumption is expressed in Coulomb or C in short.}

\subsection{Summary of guarantees offered by the ActivFORMS approach in Stage III} 

\begin{mdframed}
	\textit{Guarantees}: by estimating the qualities of the adaptation options using statistical model verification at runtime and selecting options that realize the adaptation goals with a required accuracy and confidence level, Stage III guides the adaptation of the system to realize its adaptation goals;\vspace{2pt}\\
	\textit{Scope}: the guarantees hold for the runtime models and are confined to the extent that these models capture the actual state and behavior of the managed system and its environment; further the guarantees hold for the hypothesis that are tested, hence they are confined to the extent that these hypothesis are correct. 			
\end{mdframed}
\vspace{5pt}

\section{Stage IV ActivFORMS: Evolution Adaptation Goals \& Feedback Loop}\label{subsection:stage-4}\label{section:stage_IV}

Stage IV of the ActivFORMS approach offers basic support for on-the-fly changes of the adaptation goals and the feedback loop model through the goal management layer, see Fig.~\ref{fig:activFORMS} and Fig.~\ref{fig:activFORMS-RA}. 
Support for changing the adaptation logic during operation is a key aspect of self-adaptation \cite{Cheng:2007,Lemos2013,Souza:2013,Weyns2019}, but, limited research exists in this area. On-the-fly changes of the feedback loop are important: (i) to update MAPE models and/or knowledge to resolve a problem or a bug (e.g., add or replace some functionality), and (ii) support changing adaptation goals, i.e., change or remove an existing goal, or add a new goal. Stakeholders trigger the evolution of the feedback loop model either based on feedback obtained from the executing system or because the adaptation goals need to be changed. 

Recall that ActivFORMS only supports updates of the adaptation logic that do not require updates of the managed system, including probes an effectors. 
Co-evolution of the managing system and the managed system, which remains an open problem~\cite{Weyns2019}, is out of scope of 
this paper. We focus here on adding a new adaptation goal and updating the feedback loop models accordingly. 

\subsection{Specifying and Verifying New Adaptation Goals and MAPE Models}  

\subsubsection{Specifying and Verifying New Adaptation Goals and MAPE Models with ActivFORMS}  

When stakeholders define a new requirement for adaptation, the designer needs to translate the requirement to an adaptation goal that can be processed by the MAPE models. Further, a new quality model needs to be defined that allows predicting the quality property that corresponds with the adaptation goal for the different adaptation options. These domain-specific tasks are similar to the specification of adaptation goals and quality models in Stage I. When adding a new adaptation goal, the designer also needs to update MAPE models. Usually, the monitor model needs to be extended with support to track data and uncertainties that relate to the adaptation goal. The analyzer needs to be extended with support to perform analysis of the new quality property. The planner needs to incorporate the new goal in the set of adaptation goals to select the best adaptation option. New types of plan steps may need to be incorporated in the planner, and, the executor may need to be extended to deal with new types of adaptation actions. 

To verify the correctness of the behavior of the updated MAPE models, the initial stub models for the probes, effectors, and the verifier may need to be updated to ensure that they cover the required execution paths of the MAPE models to verify the different correctness properties. For the verification, the initial correctness properties can be checked again,  possibly complemented with the verification of new domain-specific properties. 

\subsubsection{Specifying and Verifying New Adaptation Goals and MAPE Models with ActivFORMS-ta}

ActivFORMS-ta offers basic support for on-the-fly changes of the adaptation goals and MAPE models. When a new (or updated) requirement is formulated, the designer needs to specify a new adaptation goal, update the MAPE models, design new quality models and integrate these with the other feedback loop models. ActivFORMS support the designer in these tasks with its MAPE model templates, including the specification of new goals exploiting the Knowledge and adjusted MAPE models using the templates of the MAPE models. Next, the designer needs to verify the updated MAPE models to ensure that the feedback loop complies with the correctness properties. This verification typically requires an update of the stub models. The MAPE model templates, in particular the set of properties, support the designer with specifying properties. The generic properties that require an instantiation for the domain at hand ($Pr8$, $Pr9$, and $Pr11$) may also need to be extended. In ActivFORMS-ta, the Uppaal tool~\cite{tutorial04} is used to specify and verify the evolved feedback loop model.
\vspace{5pt}\\
\noindent
\textbf{Example 9.}  We show how the designer can dynamically add the latency requirement as an additional adaptation goal to DeltaIoT. Recall from Section \ref{section:iot} that DeltaIoT has a third additional requirement that needs to be activated at runtime: \textit{R3. The average latency of packets should be less than 5\% of the cycle time.} This new requirement can be translated to an adaptation goal as follows: 

\lstset{caption={Definition of adaptation goal for latency},label=q-l}
\footnotesize
{\ttfamily
	\begin{lstlisting}
   type struct {
     ... 
     int latency; 
   } Qualities
   int MAX_LATENCY = 5; 
   bool satisfactionGoalLatency(Configuration gConf, int MAX_LATENCY) {
     return gConf.qualities.latency <= MAX_LATENCY; 
   }
	\end{lstlisting}
}
\normalsize

\begin{figure}[h!tb]
	\centering
	\includegraphics[width=0.75\textwidth]{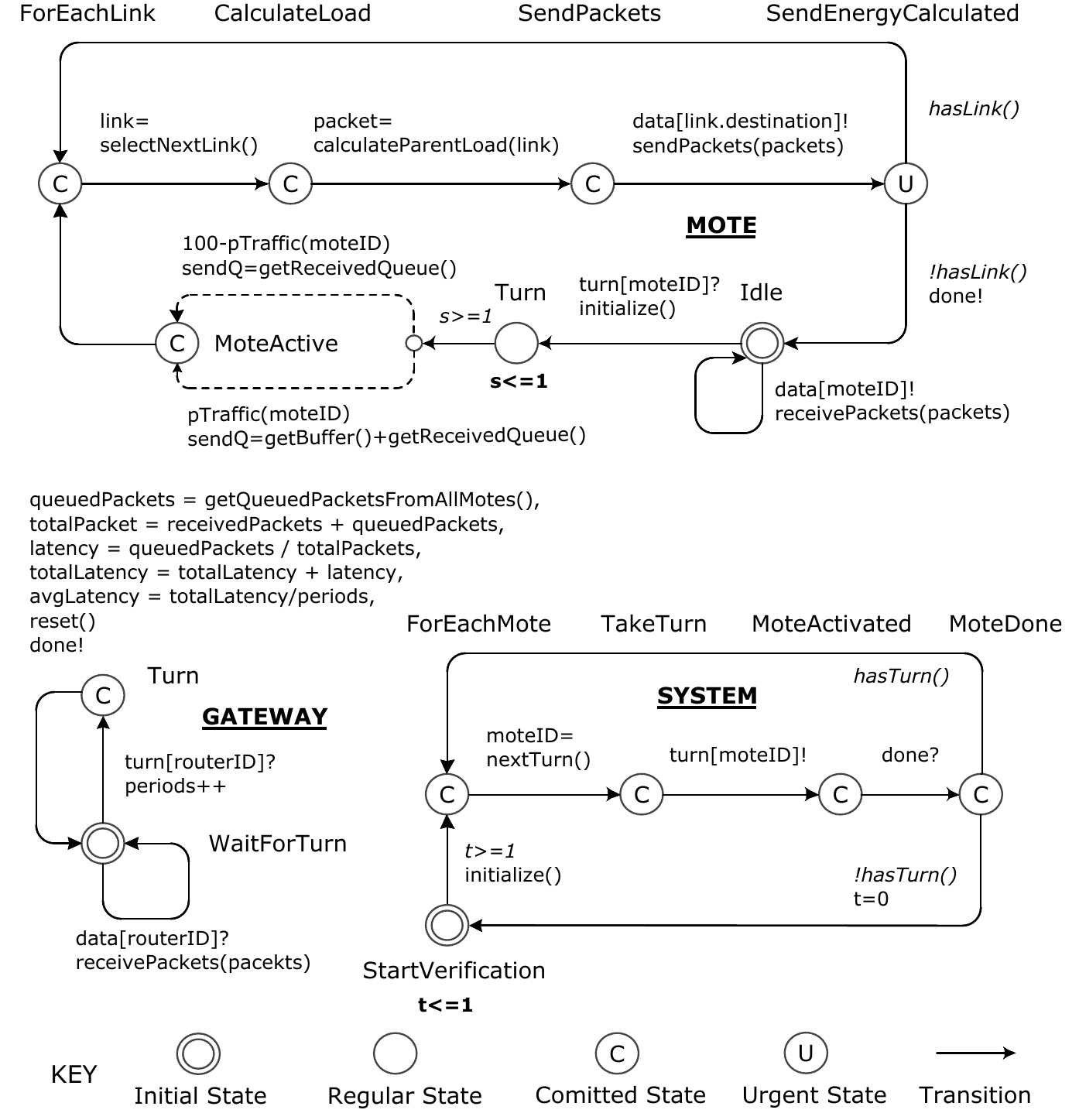}
	\caption{Latency model for DeltaIoT}
	\label{fig:latency_model}
\end{figure}

Fig.~\ref{fig:latency_model} shows the quality model to estimate latency for DeltaIoT. 

The model has a similar structure as the quality model to estimate energy consumption (available at the ActivFORMS website). For the verification of the model, the designer specified the following query: 

\lstset{caption={Verification query for latency model},label=q-l}
\footnotesize
{\ttfamily
	\begin{lstlisting}
			simulate 1[<=30](Gateway.latency)
	\end{lstlisting}
}
\normalsize

This query calculates the estimated latency for the adaptation option that is set based on 30 simulation runs (for an RSEM of $0.5\%$ determined based on offline experiments). The \textit{System} automaton activates the motes one by one ($\mathit{moteId}$ = $\mathit{nextTurn()}$). Each Mote can then send packets to its parents in its time slots ($\mathit{sendPackets(packets)}$). When the $\mathit{Gateway}$ gets its turn, it computes the latency based on the proportion of packets that did not arrive (i.e., remained in queues) compared to the total number of packets (i.e., packets ar- rived and in queues). For each simulation run the verifier assigns the uncertainty value for $\mathit{pTraffic(moteId)}$ per mote. The analyzer uses the $30$ results to compute and estimated average latency with the required accuracy.

\begin{figure}[h!tb]
	\centering
	\includegraphics[width=\textwidth]{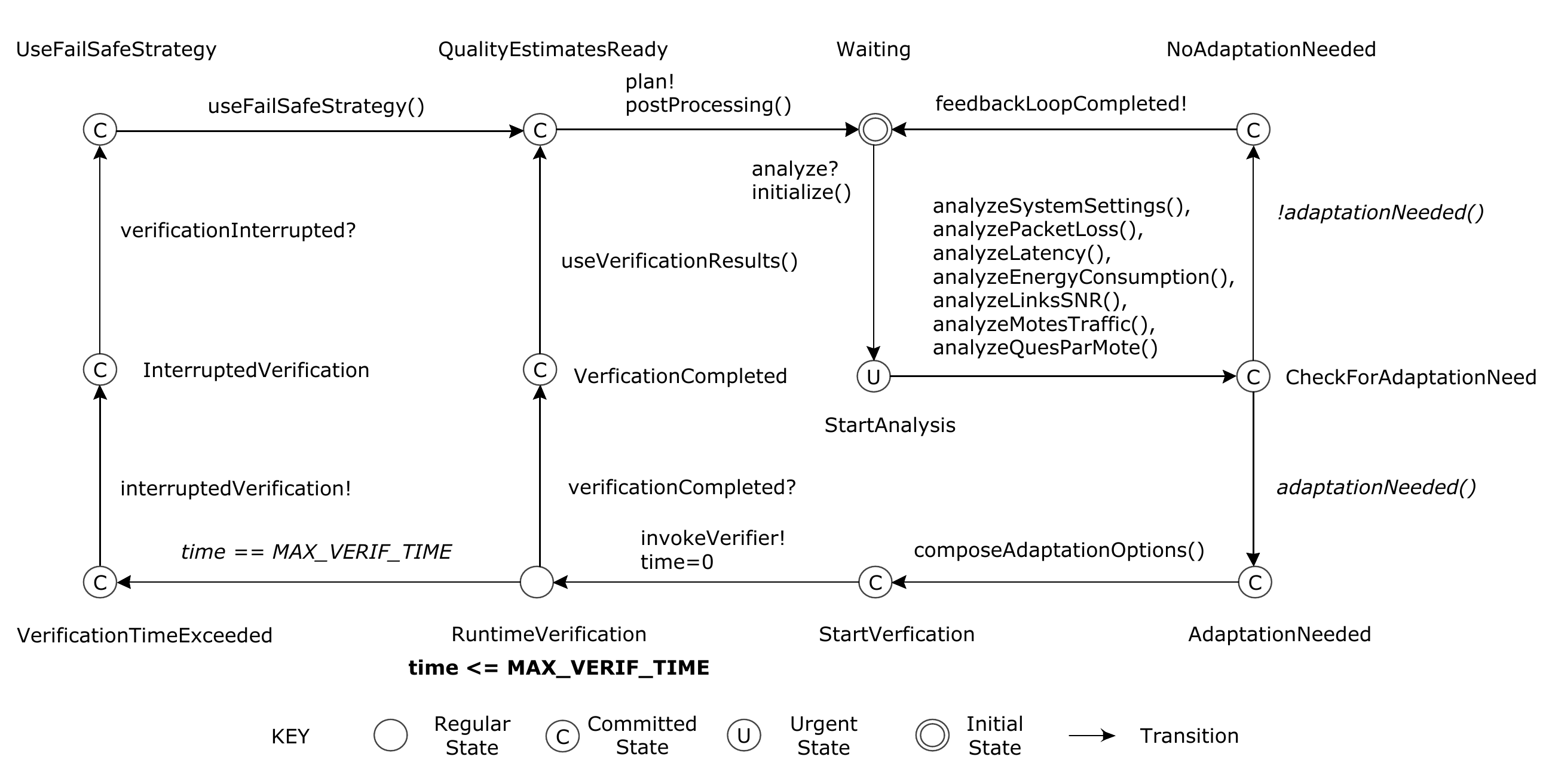}
	\caption{Updated analyzer model for DeltaIoT to deal with latency}
	\label{fig:updated_analyzer}
\end{figure}

Incorporating the new latency goal requires updating the MAPE models. Fig.~\ref{fig:updated_analyzer} shows the updated analyser model. For the other updated models and stubs, we refer to the ActvFORMS website.  

The \textit{Analyzer} model is extended with two functions: $\mathit{analyzeLatency()}$ checks whether the latency of the network is above the threshold of maximum latency,  and $\mathit{analyzeQueuesPerMote()}$ checks whether the queues for each mote are saturated or not. This data is taken into account when evaluating $\mathit{adaptationNeeded()}$. Furthermore, when the analyser invokes the verifier it will perform analysis of the latency model in addition to packet loss and energy consumption. 

For the verification of the updated MAPE models, we could reuse the initial set of properties (see Example 4). As the changes of the models were limited, the extra time required for verification time was limited. 
We discuss the verification results in the next section.

\subsection{Enact New Models} 

\subsubsection{Enact New Models with ActivFORMS} 

When the evolved feedback loop model is verified it needs to be enacted. Model enactment follows a semi-automatic process that is supported by the \textit{Goal Management} layer and the model execution engine, see Fig.\,\ref{fig:activFORMS-RA}. ActivFORMS uses the classic process to update the feedback loop model based on quiescence~\cite{Kramer:1990}. A quiescent state of a component or a model is a state where no activity is going on in the element so that it can be safely updated. The designer is responsible to determine quiescent states when specifying a concrete feedback loop model and ensure that the feedback loop model can reach these states. One approach a designer can use to realize quiescence is by means of using reactive MAPE models (we apply this in ActivFORMS-ta). In this approach, each of the MAPE models has a dedicated state where the MAPE behavior waits to be triggered to start its adaptation function. When all MAPE models are in the waiting state, the feedback loop is in a quiescent state. In addition, the goal management layer needs to be designed such that messages invoked to the feedback loop during a model update are buffered and handled after the update (e.g., messages with data from a probe), and that the state of the old model can be transferred to the new model. Handling messages that arrive during a life update of a model and transferring state from an old to the new model are difficult domain-specific problems that in general cannot be solved without human intervention\,\cite{Vandewoude2007}. In ActivFORMS, the designer is responsible to implement the message handing and state transfer functionalities. 

When a feedback loop model needs to be changed, an operator uses the \textit{Update Interface} to load the new model (see Fig.\,\ref{fig:activFORMS-RA}). This activates the \textit{Online Update Manager} to start tracking the executing feedback loop model until it reaches a quiescent state. The online update manager then triggers the model execution engine to perform the model update. To that end, the model execution engine saves the state of the old model, replaces the old model with the new model, initiates and updates the new feedback loop model, and starts the execution of the new model. To preserve the guarantees that the behavior of the new MAPE models is correct with respect to a set of correctness properties (obtained during specification and verification), the online update manager and model execution engine need to be trusted. This means that the designer of the online update manager and model execution engine of a concrete instance of ActivFORMS needs to provide the required evidence that these components perform their functions correctly with respect to loading the feedback loop model, identifying quiescent states, replacing the model, transferring the state and initializing the model, handling buffered messages, and restarting the execution of the feedback loop model. 

\subsubsection{Enact New Models with ActivFORMS-ta} 

ActivFORMS-ta supports the enactment of new adaptation goals and MAPE models through a trusted online update manager; trustworthiness is based on extensive testing. The test suite with a test report is available at the ActivFORMS website. 

Once the new feedback model is verified, an operator can use the online update manager to load the new verified model via the update interface.\footnote{ActivFORMS-ta enables an operator to load a new model from a file via a command line interface or via a graphical user interface. We refer the interested reader to the ActivFORMS website for details.} When the new model is loaded, the update manager observes the executing feedback loop model to determine when it reaches a quiescent state~\cite{Kramer:1990}. In ActivFORMS-ta, the $\mathit{Waiting}$ states of the MAPE models define the quiescent state of the feedback loop model (if one of the MAPE models is not in the waiting state, some adaptation activity going on). Once the quiescent state is reached the manager notifies the virtual machine to update the running feedback loop with the new model. The trusted update manager and virtual machine guarantee that model updates are performed consistently.
The concrete steps of model enactment are as follows:  

\begin{enumerate}
	\item The new model update is loaded via the update interface;
	\item The online update manager tracks when the MAPE models enter quiescence states;
	\item Once the models are in quiescence states the online update manager notifies the virtual machine to start updating the running feedback loop model; 
	\item The virtual machine halts the execution of the running feedback loop model and saves the state of the model; 
	\item The virtual machine adds incoming signals into a waiting queue;\footnote{These are in principle only signals from a probe that indicate the availability of new data for the monitor.}
	\item The virtual machine loads the new models; 
	\item The state of corresponding variables are copied from old to new models. New variables are initialized; 
	\item The virtual machine starts executing the new model;
	\item Pending signals waiting in the queue are processed (first-in-first-out). 
	\item Normal execution continues.  
\end{enumerate}

\begin{figure}[h!tb]
	\centering
	\includegraphics[width=0.85\textwidth]{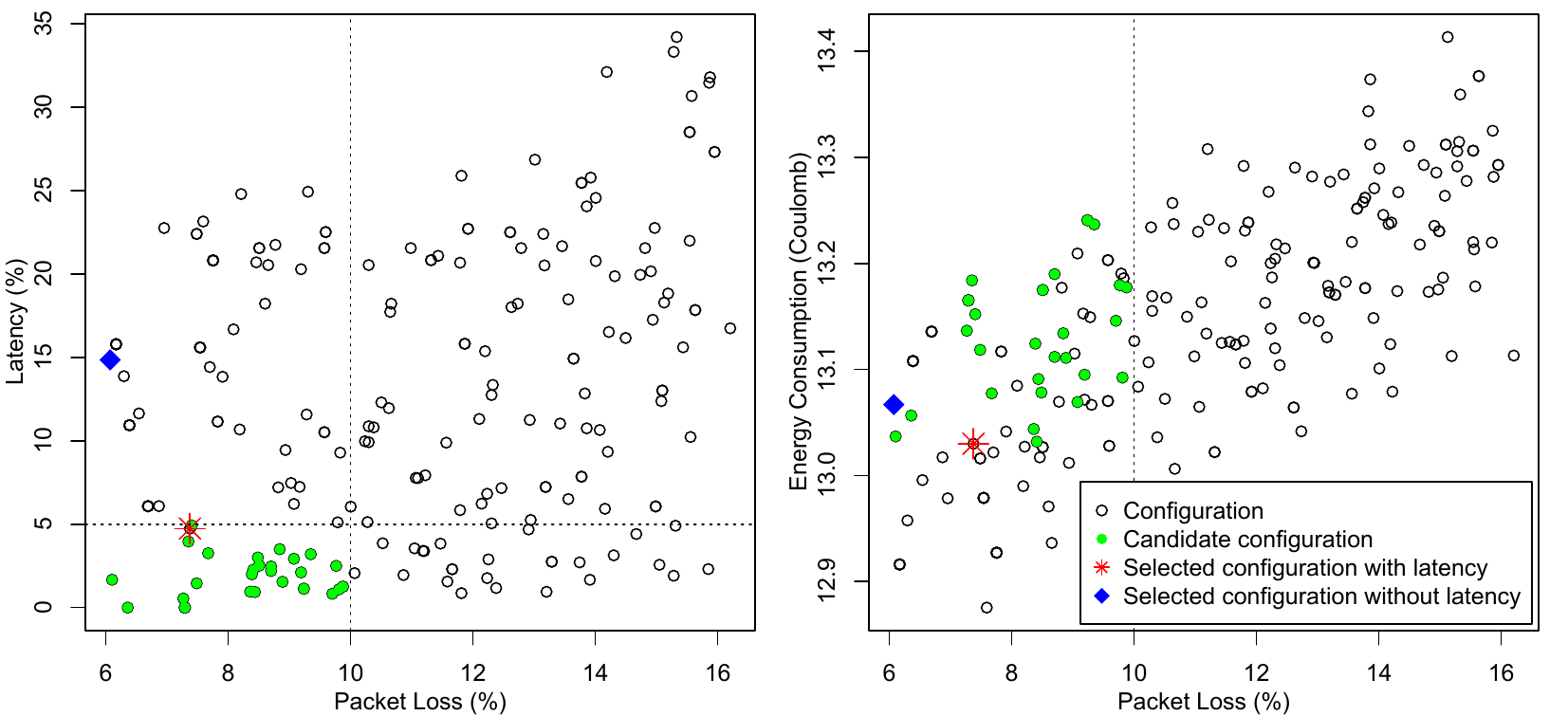}
	\caption{Selection of the best adaptation option with three adaptation goals.}
	\label{3-qualities}
\end{figure}

\noindent
\textbf{Example 10.}  We illustrates the effects of incorporating the latency goal in DeltaIoT, see Fig.~\ref{3-qualities}. 

The diagram on the left shows the adaptation options at a particular point in time. The shaded dots (marked in green plus one in red) represent adaptation options that comply with the adaptation goals for packet loss ($<10\%$) and latency ($<5\%$). The shades dots in the diagram on the right show the same adaptation options; the dot marked with a star (in red) is selected for adaptation as the expected energy consumption for this option is minimum. The selected adaptation option has an expected packet loss of $7.5\%$, a latency of $4.8\%$, and an energy consumption of $13.03$~C.
The diamond dot (in blue) in both figures shows the option that would have been selected if the latency goal would not have been taken into account (latency $14.9\%$). The figures illustrate that ensuring the latency goal may introduce a small tradeoff against the other two adaptation goals.

\subsection{Summary of guarantees offered by the ActivFORMS approach in Stage IV}

\begin{mdframed}
	\textit{Guarantees}: by evolving and verifying the feedback loop model, possible using MAPE model templates, and updating the running model with the new model using goal management and the model execution engine, stage IV guarantees the correct behavior of the model with respect to a set of correctness properties;\vspace{2pt}\\ 	
	\textit{Scope}: the guarantees hold for the updated feedback loop model under the assumptions of Stages I to III; in addition, goal management and the model execution engine should perform the model update correctly wrt. loading the new feedback loop model, identifying quiescent states, replacing the model, transferring the state, initializing the new model, handling buffered messages, and restarting model execution.
\end{mdframed}  


\section{Evaluation of ActivFORMS-ta}\label{section:evaluation}

We evaluated ActivFORMS and its tool-supported instance using the DeltaIoT network deployed at KU Leuven, shown in Fig.~\ref{fig:DeltaIoT}. The default setup consists of 15 motes, each mote comprising: (1) a Raspberry Pi that is responsible for sensing, local processing, and network management operations, and (2) a RN2483 LoRa module\footnote{http://ww1.microchip.com/downloads/en/DeviceDoc/50002346C.pdf} that is in charge of radio communication. The current consumed by the LoRa module at $3.3\,$V is $20.2\,$mA for power setting $0$ and $38.9\,$mA for power setting $15$. The gateway runs on a regular server machine that is responsible for processing network data and storing the network statistics in a database. The server offers an API to a probe and effector to monitor and adapt the network. For the simulation tests we used a Macbook with 2.5 GHz Core i7 processor, and 16 GB 1600MHz DD3 RAM. All the data of the evaluation is available at the ActivFORMS website, incl. a link to the DeltaIoT artifact that can be used to replicate the experiments.

Unless mentioned differently, we run the default setup of the DeltaIoT network with 15 motes for a period of $12$ hours. The cycle time is set to $9.5$ minutes, corresponding to $76$ cycles in $12$ hours. A cycle consists of two phases: the first $8$ minutes are allocated to the motes to communicate date downstream to the gateway; the remaining $1.5$ minutes are allocated for the communication of adaptation messages from the gateway upstream to the motes. The maximum verification time is set to $8$ minutes. Each mote can generate $10$ packets per cycle, subject to its traffic load profile, see the description in Section \ref{section:iot}. In each cycle, each mote gets $40$ slots of $2$ seconds for communication. The size of the \textit{send-queue} is $60$, which implies that packets in the queue are sent within two cycles. Packets from children that arrive when the \textit{receive-queue} is full are discarded. The values for SNR are based on the actual conditions of the wireless communication.\footnote{All other network settings are fixed during the experiment, e.g., the spreading factor of all motes is set to $8$ in all experiments. For details, we refer to the ActivFORMS website} 

We start with the offline stages of ActivFORMS-ta, where we determine the degree of reuse when applying the MAPE templates and the time required time to verify the correctness properties and this for the default setup of DeltaIoT. Then we evaluate the online stages, in particular 
the effect of the verification settings when applying runtime statistical model checking, the performance of ActivFORMS-ta compare with the runtime quantitative verification, the scalability of the approach, and the validation and effects of adding a new goal during operation. 

\subsection{Evaluation of the Offline Stages of ActivFORMS-ta} 

\subsubsection{Reuse Design Feedback Loop Model} 

To determine the degree of reuse when applying the MAPE templates, we measured the reuse of the templates for the feedback loop with the default setup of DeltaIoT.
We used a set of metrics that directly relate to the rules for instantiating the MAPE templates (see Section~\ref{subsubsec:rules_usage_templates}). We focused on the MAPE models. For each model template, we measured: (i) the number of model elements of the template that we directly reused for the instance; (ii) the number of abstractly defined elements of the template with fixed names that we implemented; (iii) the number of abstractly defined elements of the template with domain-specific names that we implemented; (iv) the number of instances we used for the facultative model constructs in the templates (name I), and (v) the number of new modeling elements.

\begin{figure}[h!tb]
	\includegraphics[width=\textwidth]{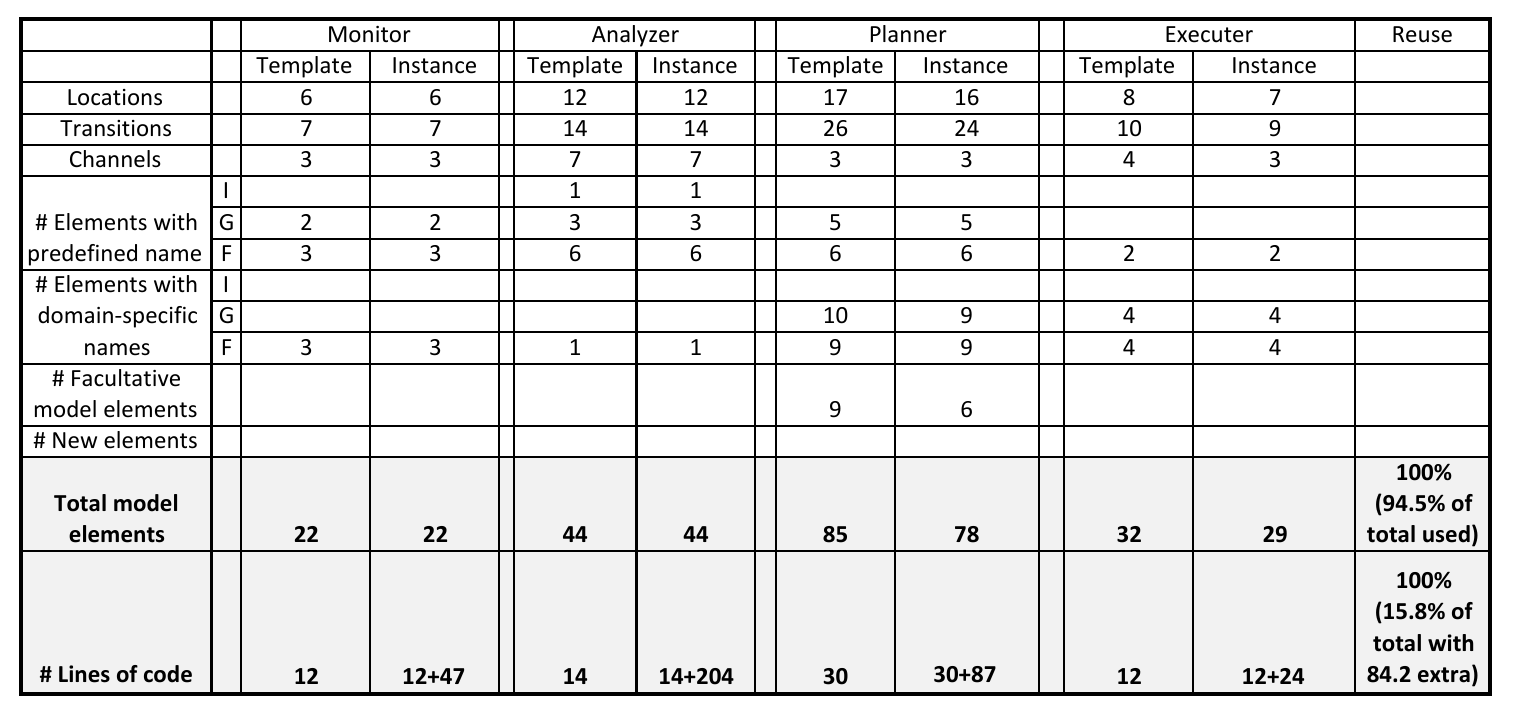}
	\caption{Modeling reuse of templates feedback loop model}
	\label{modeling_reuse_templates}
\end{figure}

\paragraph{Results} The table in Fig.~\ref{modeling_reuse_templates} shows the results. The numbers demonstrate that the templates offer a high degree of reuse for modeling elements of concrete MAPE feedback loops. Concretely, for the design of the MAPE feedback loop we reused 94.5\% of the model elements and no new modeling elements were needed. On the other hand, a substantial number of lines of code had to be added; concretely, we reused all the code of the templates that made up 15.8\% of the code; the remaining 84.2\% was new domain-specific code.

\paragraph{Conclusions} We conclude that the graphical modeling elements are highly reusable. The domain-specific instantiation is mainly realized by adding additional domain-specific code elements. Since the results apply to a single case; additional experiments are required to generalize the findings.

\subsubsection{Performance of the Verification of Feedback Loop Model}

To evaluate the performance of the offline verification of a feedback loop model; we performed experiments with the default setup of DeltaIoT as shown in Fig.~\ref{fig:DeltaIoT} and two adaptation requirements (R1. Packet loss should be less than 10\% and R2. Energy consumption should be minimized). The detailed models and stub data that we used for the verification with the Uppaal model checker are available at the ActivFORMS website.

\begin{figure}[h!tb]
	\centering
	\includegraphics[width=0.75\textwidth]{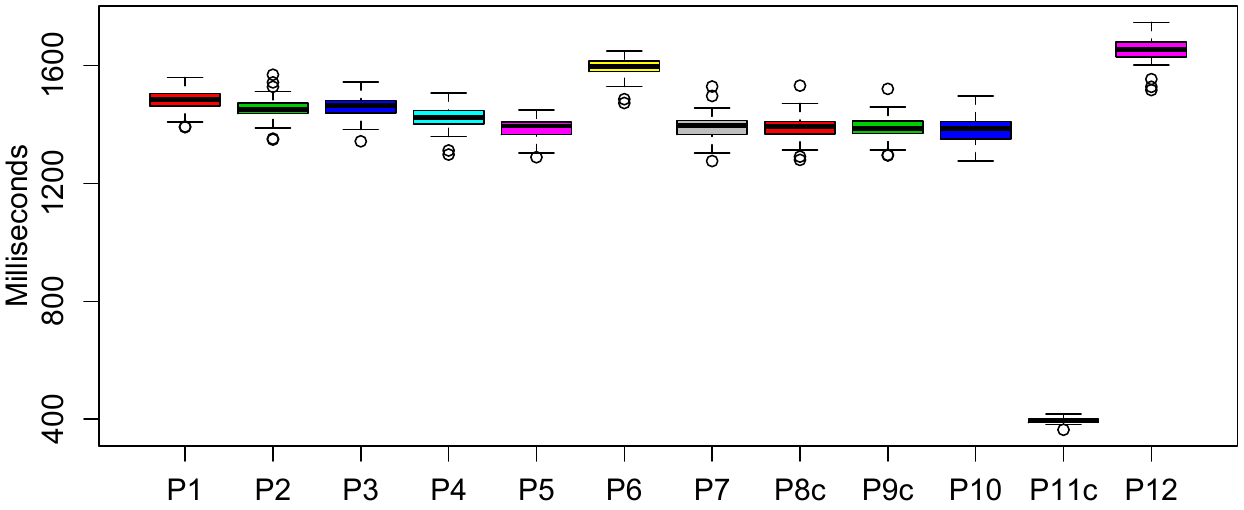}
	\caption{Verification times for properties that check the correctness of the MAPE feedback loop}
	\label{design_time_verification_results}
\end{figure}

\paragraph{Results}
Fig.~\ref{design_time_verification_results} shows the time taken to verify the MAPE feedback loop properties (average results of 50 runs). The number of states explored for the different properties varied between 731 and 3158, with an average of 2630. When incorporating the latency goal (Stage IV), the verification time increased with $55.1\%$ to an average of $2.12$\,sec. A detailed report of these verification results is available at the ActivFORMS website. 

\paragraph{Conclusions} 
The results show that the overhead for design-time verification of the properties that check the correct behavior of the MAPE feedback loop with respect to the correctness properties provided by the templates is totally acceptable for a default setup of DeltaIoT, which is a realistic IoT application setting (average $1.55$\,sec).

\subsection{Evaluation of the Online Stages of ActivFORMS-ta}

The evaluation of the online stages aims at answering the following 
three research questions:

\begin{itemize}
	
	
	\item[] RQ1: How does ActivFORMS-ta with statistical model checking compare with a reference approach and runtime quantitative verification for achieving the quality goals of the setup of the DeltaIoT network? \vspace{3pt}
	
	\item[] RQ2: How does adaptation time and memory usage for ActivFORMS-ta with statistical model checking compare with the runtime quantitative verification for setups of DeltaIoT with increasing scale? \vspace{3pt}
	
	\item[] RQ3: What is the impact on the quality goals and adaptation time for ActivFORMS-ta when a new adaptation goal is dynamically added to the setup of the running DeltaIoT network?
	
\end{itemize}

Section~\ref{performance} answers RQ1 by comparing ActivFORMS-ta with an over-provisioning approach that is commonly used in practice (referred as the reference approach) and a state-of-the-art approach that uses runtime quantitative verification to estimate the qualities of the adaptation options (RQV, e.g., used in~\cite{Calinescu2011,DBLP:conf/sigsoft/MorenoCGS15,DBLP:conf/icsa/CamaraMV20}). In Section~\ref{scalability}, we answer RQ2 by comparing the scalability of ActivFORMS-ta with RQV. We measure the adaptation time and memory usage for network configurations with an increasing number of motes. Finally, in Section~\ref{adding_goal}, we answer RQ3. We dynamically incorporate the latency goal in the network with ActivFORMS-ta and test the impact of it.

\subsubsection{Comparison ActivFORMS-ta with Reference Approach and RQV}\label{performance}

The experiments in this part are done on the real deployment of the IoT network. We consider two adaptation goals: R1. Average packet loss should be below 10\% and R2. Energy consumption should be minimized. For R1, we used an approximation interval with $\varepsilon$\,=\,$0.01$, i.e., $E$\,=\,$99\%$ accuracy, and confidence $\alpha$\,=\,$0.10$, i.e., $A$\,=\,$90\%$. For R2, we used an RSEM of $0.5\%$, which requires 30 simulation runs (see  Section~\ref{subsubsection_analysis_options}). We compare ActivFORMS-ta with a reference approach where all motes communicate at maximum power and send all packets to all their parents. This over-provisioning approach is commonly used in practice to assure high packet delivery performance at the cost of reduced lifetime of the network. We also compare ActivFORMS-ta with a state-of-the-art analysis approach that uses RQV. For RQV, we translated the automata models for the quality properties to a Discrete Time Markov Chain model for packet loss and a Markov Decision Process model for energy consumption. For the verification at runtime we enabled the analyzer model to use the PRISM model checker \cite{KNP11} with the default settings. The definition of the RQV models is available at the ActivFORMS website. For the verification with runtime SMC and RQV we verify the adaptation options sequentially. 
In every step, an adaptation option is randomly selected and the quality properties of the option are verified one by one, first packet loss and then energy consumption. The option is then removed from the list and the process is repeated until all the adaptation options are verified or the maximum verification time is reached.

\begin{figure}[h!tb]
	\centering
	\includegraphics[width=0.9\textwidth]{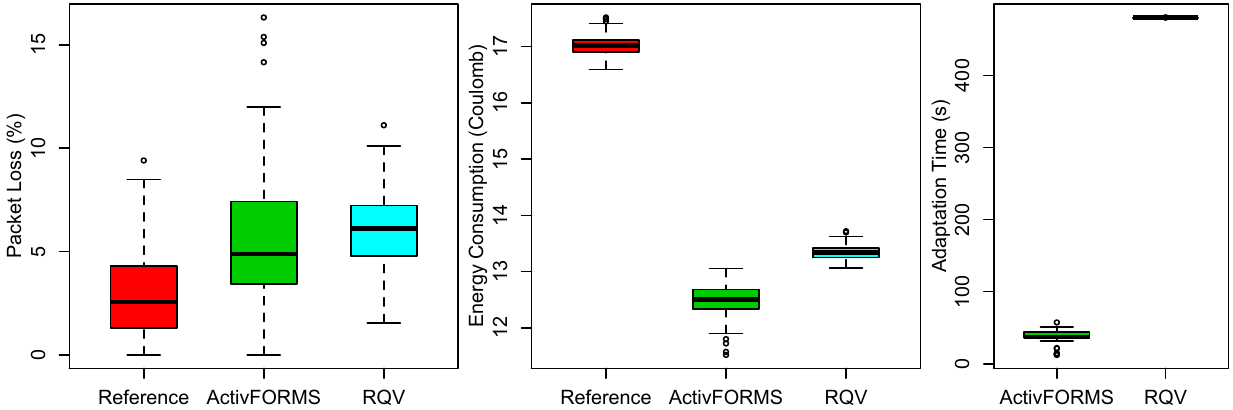}
	\caption{Results for a DeltaIoT setting with 15 motes and two adaptation goals.}
	\label{fig:results}
\end{figure}

\paragraph{Results}
Fig.~\ref{fig:results} shows the results for runs of $12$ hours with adaptation applied every cycle of $9.5$ minutes. For the verification of quality models with ActivFORMS the maximum number of explored states was 41.8K and with RQV 17.1K. 
The reference approach realizes $R1$ with a better result for packet loss as ActivFORMS-ta (mean $2.6\%$ versus $4.9\%$ for  ActivFORMS-ta). 
However, ActivFOMRSi significant reduces energy consumption with about 27\% ($12.5$~C compared to $17.2$ for the reference approach). RQV on the other hand realises a slightly worst result as ActivFORMS-ta on packet loss (mean $6.1\%$) and energy consumption (mean $13.3$~C). Although RQV applies exhaustive verification, due to time constraints ($\leq8$\,min verification time), the approach verified only a fraction of the adaptation options and hence was not able to find the best solution. Concretely, for the IoT setting with 15 motes and 2 requirements, on average, RQV was able to complete the verification of only $8\pm1$ of the adaptation options on a total of $216$ options within the verification time period.

Fig.~\ref{fig:results} on the right shows the adaptation times (primarily used for verification). On average, ActivFORMS-ta used $37$\,sec to compute the verification results and realize adaptation. RQV used the complete available time slot, resulting in an average adaptation time of $480$\,sec, i.e., $8$\,min.\footnote{Analysis was interrupted when the verification of the last option that was started within 8 min completed.} 

\paragraph{Conclusions} 
In answer to RQ1, compared to the reference approach, ActivFORMS-ta realizes the packet loss requirement with sufficient accuracy and confidence, while it significantly improves energy consumption with about $27\%$ for a default setup of DeltaIoT. ActivFORMS-ta realizes self-adaptation in a time window that is only a fraction of the cycle time of $9.5$ minutes. For this setup, RQV could verify only a fraction of the adaptation options within the given time, it realizes slightly worst results for both requirements compared to ActivFORMS-ta with statistical model checking.

\subsubsection{Scalability of ActivFORMS-ta Compared with RQV}\label{scalability}

To evaluate the scalability of ActivFORMS-ta (with SMC) and compare it with RQV, we measured the adaptation time and the memory usage for network settings with increasing complexity. We performed these tests in simulation. Concretely we increased the number of motes of the IoT network from 5 to 25 in steps of 5.\footnote{To ensure a fair comparison between ActivFORMS-ta and RQV, the randomly selected adaptation options and uncertainty settings were recorded from the tests with ActivFORMS-ta and the same settings were used for the tests with RQV.} The number of adaptation options in the network is defined by $6^{\frac{m}{5}}$ with $m$ the number of motes; e.g., a setting with 10 motes has 36 options, while a setting with 25 mote has has $7776$ options. We applied adaptation for packet loss (accuracy $E$\,=\,$99\%$ and confidence $A$\,=\,$90\%$) and energy consumption (RSEM $0.5\%$). All material of the experiments is available at the ActivFORMS website.

\begin{figure}[h!tb]
	\centering
	\includegraphics[width=\textwidth]{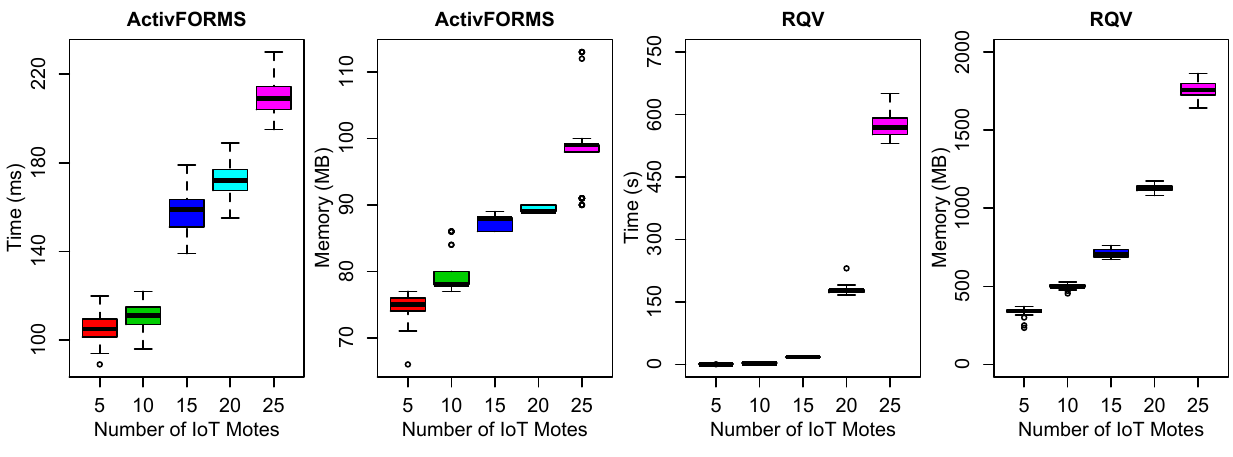}   
	\caption{Results of the scalability tests for ActivFORMS and RQV (based on 100 runs for one randomly selected adaptation option per network configuration with 5, 10, 15, 20, and 25 motes respectively)}
	\label{fig:latency_results}
\end{figure}

\paragraph{Results}
Figure~\ref{fig:latency_results} shows the results for ActivFORMS-ta on the left hand side and RQV on the right hand side. The graphs show the results of 100 runs, each based on a randomly selected adaptation option for IoT networks with 5 to 25 motes. The results show that ActivFORMS-ta scales well for networks up to 20 motes, both for verification time and memory usage. If we multiply the mean verification time of $168$\,ms for a setting with 20 motes that has $1296$ adaptation options, the total verification time would be $3.63$\,m, which is less than half of the available 8 m. For the configuration with 25 motes with $7776$ adaptation options, the total verification time would be $26.44$\,m. However, if we relax the verification settings slightly to accuracy $E$\,=\,$95\%$, confidence $A$\,=\,$90\%$, and RSEM\,=\,$1\%$, the verification time with ActivFORMS-ta decreases to $6.48$\,m and the solution would scale well, although with slightly less accurate verification results. ActivFORMS-ta requires $75$ to $100\,$MB memory for verification. RQV does not scale for more complex networks. Multiplying the mean verification time of $17.45\,s$ for 15 motes with $216$ adaptation options, would require around $62.82$\,m, i.e., $8$ times the available time of $8$\,m. In addition, RQV requires $500$ to $1800\,$MB for verification. 

\paragraph{Conclusions}
In answer to RQ2, the test results show that ActivFORMS-ta scales well for IoT networks with $25$ motes and up to $10\,$K adaptation options. With RQV complete verification is limited to settings with $10$ motes, and requires up to $20$ times more memory as ActivFORMS-ta. 

\subsubsection{Dynamically Incorporating Latency Goal}\label{adding_goal}

To conclude, we dynamically add a latency goal to the running system and evaluate the impact of it on the quality properties and the adaptation time (Stage IV). We used the same physical setup as in part three with $15$ motes and accuracy $E$\,=\,$99\%$ (i.e., $\varepsilon$\,=\,$0.01$) and confidence $A$\,=\,$90\%$ ($\alpha$\,=\,$0.10$). We started with the packet loss and energy consumption goals only. After 12 hours, the latency goal was dynamically added for 12 hours. The latency model and the approach to add the goal dynamically is explained in Examples 8 and 9.  

\paragraph{Results}
Figure~\ref{fig:latency_results} shows the test results. As we can see, adding the latency goal drastically reduces the latency. We could not measure any latency ($0.00\,\%\pm\%0.00$) compared to $18.60\,\%$ and -$7.50$/+$7.00$ for the setup without latency goal. This improvement has only a small effect on packet loss (mean from $5.78\,\%$ to $7.10\,\%$) and energy consumption (mean from $12.47$ to $12.72$~C). On the other hand, verifying the latency model increased the overall verification time from with $24.29\,$sec to $46.22\,$sec. \vspace{-5pt}

\begin{figure}[h!tb]
	\includegraphics[width=1\textwidth]{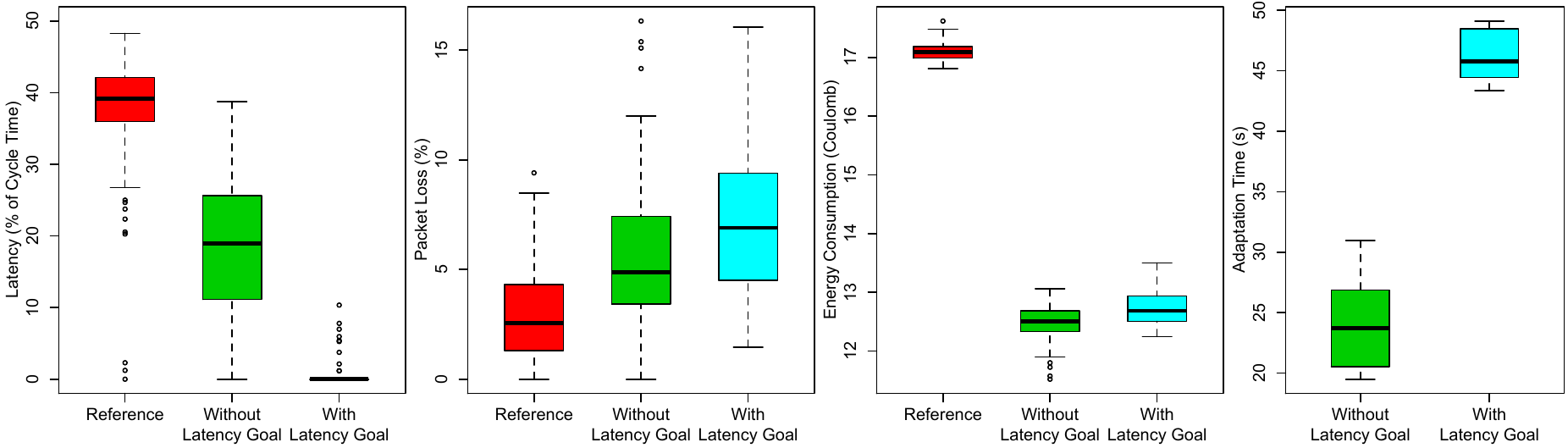}
	\caption{Impact of dynamically adding a latency goal}\vspace{-10pt}	
	\label{fig:latency}
\end{figure}

\paragraph{Conclusions}
In answer to RQ3, the test confirms that ActivFORMS-ta supports adding a new goal during operation. The results show that adding the goal drastically reduces latency of packet delivery at the cost of a small increase in packet loss and energy consumption. However, about two times more time is required for the verification of the quality models with the newly added goal.

\subsection{Threats to Validity and Limitations} 

We conclude this section with a discussion of validity threats of the evaluation of the ActivFORMS instance. Then we summarize the limitations of ActivFORMS and its concrete instance. 

\subsubsection{Threats to Validity} We discuss internal validity, external validity, reliability, and construct validity of the research results obtained for ActivFORMS-ta.

\textit{Internal validity} is about the extent to
which a causal conclusion based on our study is warranted. A part of the experiments with DeltaIoT were done in a real-world setting, hence the system may have been exposed to various external influences. Consequently, variables other than the independent variables may have act on the system at the same time. To avoid this threat, we have performed extensive trial tests before performing the experiments to ensure proper experimental conditions, including determining proper locations of the motes and defining the setting of fixes network parameters. For the experiments in simulation, we emulated the network environment based on experimental data obtained from extensive field tests.  

\textit{External validity} is about generalizing our findings to other instances of ActivFORMS and application domains. We have evaluated one concrete tool-supported instance of ActivFORMS and applied it in this paper to one application domain. Consequently, the claims we make about the usability and efficiency to engineer self-adaptive systems with ActivFORMS is only based on this instantiation. However, as we have applied several of the underlying principles of ActivFORMS to other domains, there is ground to believe that the outcomes of this paper may be generalizable. Nevertheless, we acknowledge that more research and experimentation is required to further generalize the findings obtained from the research presented in this paper.  

\textit{Reliability} is about ensuring that our results are the same if our study would be conducted again. Lack of clarity  about the research setting and the material used for performing the experiments may cause a treat to reliability. We anticipated this treat by making all the study material, data of the experimental setting, and study results publicly available. The IoT network and its simulator that we used for the evaluation are publicly available, so the results can be reproduced. Another important aspect of reliability concerns how adequate the stochastic semantics of the language used to specify timed automata in this paper are to capture the environmental uncertainty. This may need a more in depth analysis, which is beyond the scope of this paper. 

\textit{Construct validity} is about the degree to which our tests measure what they claim to measure. From the reported test results, we derive conclusions about the correctness, efficiency, and scalability of the ActivFORMS instance for the IoT application, compared to other approaches. To determine these properties, we relied on well-established metrics. To compare ActivFORMS-ta with RQV, we had to translate automata models into Markovian models. To ensure that these models are appropriate, the models were crosschecked by an expert in Markovian models and the PRISM tool.

\subsubsection{Assumptions and Limitations}  

ActivFORMS offers a novel approach to engineer self-adaptive software systems, tackling a number of open challenges in the field. However, the approach is not generally applicable. The main limitations of ActivFORMS in general and its concrete instance ActivFORMS-ta in particular are: 
%
\begin{itemize}
	\item[$-$] The managed system needs to be available and instrumented with probes and effectors; 
	\item[$-$] The managed system needs to provide support to realize consistent adaptations (changing parameter settings, adding/replacing/removing elements, etc.); 
	\item[$-$] Engineers need access to domain knowledge to devise a feedback loop for the problem at hand; 
	\item[$-$] The necessary stub models can be devised to verify of the behavior of the feedback loop model with respect to a set of correctness properties;  
	\item[$-$] Feedback loop models can be devised that are verifiable and executable; 
	\item[$-$] Dynamics are such that the feedback loop has sufficient time to make adaptation decision;
	\item[$-$] The quality properties that are subject to adaptation can be verified using using statistical model checking (which implies that distributions of variables that represent uncertainties are known and that bounds on accuracy and confidence for the adaptation goals can be defined); 
	\item[$-$] A failsafe strategy can be specified and the feedback loop model can obtain the required knowledge about when and how to apply the strategy; 
\end{itemize}

\noindent The current implementation of ActivFORMS-ta has the following limitations: 
\begin{itemize}	
	\item[$-$] The set of adaptation options is bounded and limited (possibly by discretising the settings of actuators with continuous domains); 
	\item[$-$] Distinct adaptation goals can be defined to select the best adaptation option; 
	\item[$-$] The number of runs that are required to obtain a required accuracy for a simulation query can be obtained offline; 	
	\item[$-$] Uppaal puts restrictions on the expressiveness of model and property specifications, and the way non-determinism and untimed choice of actions is treated by the sampling procedure; 
	\item[$-$] ActivFORMS-ta does not support adaptive systems with properties that involve rare events; 
	\item[$-$] Dynamic updates of the feedback loop model requires that the feedback loop model can reach quiescent states and can be consistently updated.
\end{itemize}

\section{Related Work}\label{section:related-work}

A vast body of work exists on techniques for guarantees in self-adaptive systems, for overviews see~\cite{Camara2013,Cheng2014,Lemos2017,Weyns:2016}.  
Aligned with the presented research, we focus on approaches that use formal techniques to provide guarantees. We have structured related work in three groups: approaches that provide guarantees at design time, runtime approaches,  and hybrid approaches. For each group, we discuss a selection of representative approaches. To conclude, we position ActivFORMS in the current landscape of research and we summarize our initial work ActivFORMS leverages. 
%
%

\subsection{Design Time Approaches}  
\cite{Zhang2006} use Petri Nets to model adaptive systems; the models are automatically translated to executable programs. Properties specified in linear temporal logic (LTL) allow verifying invariants and constraints about the system and its goals, e.g., 
``the adaptive program should tolerate 2 packet loss throughout its execution.'' Conformance between the models and programs is guaranteed using model-based testing. 
\cite{Autili2015} deal with partial knowledge by automatically producing service-oriented systems in two phases. The first phase (elicit) applies a technique called StrawBerry that takes service descriptions to derive behaviour automata of the service interactions. The second phase (integrate) takes the automata to automatically synthesise a service choreography that satisfies the system goal. The approach relies on tools to guarantee functional correctness-by-construction.
\cite{Camara2017} propose an approach for evaluating the resilience of self-adaptive systems by applying robustness testing techniques on the controller to uncover failures that can affect system resilience. The approach, that is based on probabilistic model checking, quantifies the probability of satisfaction of system properties when the target system is subject to controller failures. The responses to malformed input between controller and target system are used to classify robustness. 


In summary, the related approaches in this group provide in essence guarantees based on the principle of correctness-by-construction. Consequently, the guarantees are based on the knowledge available at design time. With ActivFORMS, correctness-by-construction is applied at design time to provide guarantees for the correctness of the behavior of the feedback loop model with respect to a set of properties. These guarantees are complemented with guarantees for quality goals obtained during operation based on data of uncertainties collected at runtime. 

\subsection{Runtime Approaches} 

\cite{Calinescu2011} use a probabilistic model of an adaptive system and apply RQV to identify optimal system configurations under changing conditions. Adaptation goals are expressed as probabilistic temporal logic formulae. The MAPE components exploit different tools to assure the quality goals. Techniques, such as caching and lookahead, can be used to improve the efficiency of RQV. 
\cite{Malek2011} present POISED, a quantitative approach for making adaptation decisions under uncertainty. POISED builds on possibility theory (that is grounded in fuzzy mathematics) to assess both the consequences of uncertainty. At runtime, POISED makes adaptation decisions, i.e., runtime reconfigurations of its customisable software components, that result in the best range of potential behaviour, improving the system's quality of service. 
\cite{Filieri:2014} propose an approach that relies on control theory. The approach automatically learns a system model and synthesises a PI controller at runtime, providing control-theoretic guarantees for stability, overshoot, setting time and robustness of the system operating under disturbances, and this for one setpoint goal. A Kalman filter and a change point detection mechanism enable updating the system model on-the-fly.  
\cite{Moreno:2015} apply proactive adaptation under uncertainty. The approach uses a probabilistic model of the adaptive system in which the adaptation decision is left underspecified through nondeterminism. At runtime, a probabilistic model checker resolves the nondeterministic choices so that the accumulated utility over a horizon is maximised, taking into account the inherent uncertainty of the environment predictions.
\cite{Su2016} propose Iterative Decision-Making Scheme (IDMS) that infers point and interval estimates of transition probabilities in a Markov Decision Process (MDP) using runtime data. IDMS iteratively computes a confidently optimal scheduler that minimizes the cumulative cost for a given reachability problem, and is flexible to adjust the criterion of confident optimality and the sample size within the iteration.
\cite{Camara2016} describe an approach that applies stochastic games between two players: the adaptive system and the environment. By considering minimum and maximum rewards of the system player, independently of the strategy followed by the environment, the approach is able to identify the best and worst case adaptation scenarios with and without latency, using probabilistic model checking. 
\cite{Su2017} present Proactive performance Evaluation (ProEva) that relies on Continuous-time Markov chains (CTMCs) to analyze time-bounded performance metrics. The approach addresses the problem of providing accurate model parameters of CTMCs at runtime. ProEva extends the conventional technique of time-bounded CTMC model checking by admitting imprecise, interval-valued estimates for transition rates. ProEva computes asymptotic expressions and bounds for the imprecise model checking output. 
\cite{DBLP:journals/jss/CamaraGS19} propose an approach for proactive self-adaptation that combines machine learning with probabilistic model checking. Machine learning is used to select the best adaptation pattern for a given scenario, and quantitative verification checks the feasibility of the adaptation decision, and provides feedback to the learner helping to achieve faster convergence towards optimal decisions.


In summary, the main focus of related work in this group is on guarantees for adaptive systems based on knowledge obtained during execution. Existing work primarily relies on exhaustive verification, which is time and resources demanding. A number of related approaches relax the conditions, e.g., by focusing on specific cases or by admitting imprecision in the models and verification. ActivFORMS on the other hand relies on statistical model checking to provides guarantees for quality goals at runtime, which is more efficient, but with inherent bounds on  accuracy and confidence. In addition, ActivFORMS also provides guarantees for the functional correctness of the feedback loop and supports on-the-fly updates of adaptation goals. 

\subsection{Hybrid Approaches} 

FLAGS \cite{Baresi:2010} is a goal-driven approach for self-adaptation that spans design and runtime. The approach supports modelling both crisp goals specified in linear temporal logic and fuzzy goals specified in fuzzy temporal language. These models can then used at runtime to monitor goal violations that trigger a modification of the goal model to enforce adaptation on the running system. Related approaches are RELAX  \cite{Whittle2009}, a textual language for specifying requirements with first-class support for uncertainty, and ``awareness requirements'' that describe the situations that require adaptation and ``evolution requirements'' that prescribe what to do in these situations~\cite{Souza:2013} . 
EUREMA \cite{Vogel:2014} offers a domain-specific language to model feedback loops and their interactions. At design time feedback loop models are specified by means of operations, runtime models, and interactions. An additional layer diagram specifies the interactions between the feedback loops and the managed system. At runtime, EUREMA offers an interpreter that directly interprets the models to realise adaptation. Additionally, the models can be dynamically adjusted, supporting evolution. 
\cite{Nahabedian:2016} present a general approach to specify correctness criteria for the dynamic update of a system and a technique to automatically compute a controller that handles the transition from the old to a new specification. The approach syntheses a controller guides the system to a safe state in which the update can start, ensuring that the update will eventually occur and satisfy the new specification.
\cite{Filieri:2016} offer a mathematical framework for efficient run-time decision-making. At design time a pre-computation is applied taking a model of the system and desired goals to generate a partially evaluated set of symbolic expressions that represent verification conditions to be satisfied to meet the goals. At runtime, the actual values are bound to the variables enabling the expressions to be evaluated efficiently.  The focus of the work is on quality goals, such as reliability or energy consumption.
\cite{Cailliau:2017} apply risk analysis whereby obstacles to system goals are identified, assessed, and resolved through countermeasures. During design, obstacle/goal trees are specified together with predicates that determine the satisfaction rates of probabilistic goals. At runtime, the system is monitored and when goals are not satisfied, alternative countermeasures are selected and the goal model is updated. The running system is then adapted according to the selected countermeasures. 


The related approaches in this group combine formal techniques at design time and runtime to provide guarantees for the adaptive system. In addition to the efficiency of decision-making at runtime, ActivFORMS integrates design time and runtime guarantees with first-class support for dealing with changing adaptation goals and updating runtime models. The most closely related approach is EUREMA \cite{Vogel:2014}. However, this approach has no formal basis and consequently cannot provide the guarantees that ActivFORMS can give.

\subsection{Position of ActivFORMS in the Current Landscape}
The use of formal techniques in self-adaptive systems has gained increasing attention in recent years~\cite{Tamura2014,Lemos2017,Weyns:2016}. We contrast ActivFORMS with existing work. First, existing work offers guarantees for adaptive systems based on the principle of correctness-by-construction. The focus is primarily on a correct transition of the managed system. ActivFORMS provides fine-grained guarantees for the correct behavior of the feedback loop with respect to a set of correctness properties by: (i) formal modeling and verification exploiting reusable design knowledge in the form of formal templates, and (ii) direct execution of the verified model using a trustworthy model execution engine. Second, existing work claims to offer guarantees for the adaptation goals of the adaptive system using resources intensive exhaustive verification techniques. ActivFORMS provides guarantees that adaptation options are selected that guide the system towards its adaptation goals and this selection is done in an efficient way by using runtime statistical verification. There is a tradeoff between accuracy and confidence on the one hand and the time required for verification on the other hand, but the approach allows to set this tradeoff as required. In our work, we complement the guarantees provided by  verification at runtime to select adaptation options with evidence through validation to demonstrate that the system complies with its requirements. Third, while multiple researchers argue for the importance of uncertainty with respect to changing adaptation goals, little work exist in this area. ActivFORMS provides first-class basic support for changing the adaptation goals and the feedback loop on-the-fly relying on a trustworthy update infrastructure. In summary, existing work typically focuses on particular stages of engineering self-adaptive systems. ActivFORMS on the other hand, offers a end-to-end approach that spans the four main stages of adaptive systems, providing guarantees for the different aspects of these systems, subject to a set of assumptions.
%

To conclude, we summarize our initial related work and contrast it with the main contributions: ActivFORMS and its instance ActivFORMS-ta, see Table \ref{iw-summary}.

\begin{table}[h!]\caption{Summary of initial work and comparison with ActivFORMS and ActivFORMS-ta contributions}\label{iw-summary}
	\centering
	\begin{scriptsize}
		\renewcommand{\arraystretch}{1.2}
		\setlength{\tabcolsep}{0.4em}
		\begin{tabular}{p{2.1cm}p{2.1cm}p{2.1cm}p{2.1cm}p{2.1cm}p{2.1cm}}
			\Xhline{0.6pt}
			\textbf{Initial work}   
			&  \textbf{Model templates}
			&  \textbf{Executable models}
			&  \textbf{Decision-making}
			&  \textbf{Changing goals}
			&  \textbf{Validation}
			\\ \Xhline{0.6pt}
			SEAMS 2014 \cite{Iftikhar2014}  
			&  
			& Initial realization of virtual machine for executable models
			&              
			& Initial support for online updates of adaptation goals  
			& Simple robotic case                                                                                                                                                           \\ 	\Xhline{0.4pt}
			TAAS 2015~\cite{Didac2015}  
			& Initial set of templates used for design-time model checking 
			& 
			&     
			& & Illustrated with simulated mobile learning case and robotic case                                                   \\ 	\Xhline{0.4pt}
			M\@\,RT 2016~\cite{7573167}  
			& 
			&  
			& Study of the use of runtime simulation  
			& & Simulated service-based system
			\\ 	\Xhline{0.4pt}
			TSE 2017~\cite{Calinescu2017}  
			& Application of initial set of templates~\cite{Didac2015}  
			& Application of initial realization of executable models~\cite{Iftikhar2014} 
			&     
			& &  Simulated trading system and simulated unmanned underwater vehicle system
			\\ 	\Xhline{0.6pt}
			ActivFORMS \& \newline ActivFORMS-ta   
			& Advanced set of MAPE model templates with set of instantiation rules 
			& Trusted virtual machine for executing networks of timed automata 
			& Statistical model checking at runtime to verify qualities and threshold and optimization goals for decision-making
			& Trusted live update manager with basic support for online updates of adaptation goals and feedback loop model
			& Real-world IoT network deployment complemented with simulation for scalability tests
			%
			\\ \Xhline{0.6pt}
		\end{tabular}
	\end{scriptsize}
\end{table}

\section{Conclusions and Future Work}\label{section:conclusions}

Engineering trustworthy self-adaptive system is challenging since these systems need to handle uncertainties at runtime with time and resource constraints. To tackle this challenge, we presented ActivFORMS, a reusable model-driven approach for engineering self-adaptive systems, along with ActivFORMS-ta, a concrete instance. The  approach provides: 1) correctness of the feedback loop model with respect to a set of correctness properties that are preserved by direct execution of formally verified model using a reusable virtual machine, 2) efficient guarantees that adaptation options are selected that guide the system to realize its adaptation goals with a required level of accuracy and confidence using statistical model checking at runtime, and 3) basic support for on-the-fly changing adaptation goals and updating of verified feedback loop models that meet the new goals. ActivFORMS-ta comes with a set of reusable model templates to specify and verify MAPE-based feedback loops, it offers a trusted virtual machine to execute the models to realize adaptation, it applies runtime statistical model checking to make efficient adaptation decisions, and it offers basic support for on-the-fly updates of adaptation goals and feedback loop models using a trusted online update manager. We successfully validated ActivFORMS-ta for a real-world IoT application. 

In our future work, we plan to study how we can apply online learning techniques to deal with large spaces of adaptation options, for initial results we refer to\,\cite{QuinWBBM19}. A key challenge will be to define the impact on the guarantees that can be obtained. We also plan to study how ActivFORMS can be applied to adaptation problems with more complex types of uncertainties, such as uncertainties of the structure of models. Finally, we plan to study how ActivFORMS can be applied in systems that require multiple feedback loops that need to work together to solve an adaptation problem. 

\section{Acknowledgments}
We are grateful to VersaSense, Gowri Sankar Ramachandran, and Ritesh Kumar Singh for setting up DeltaIoT. We express our appreciation to Prof. Danny Hughes and his team at DistriNet for their continuous support in our research. We also appreciate the feedback we received from Axel Legay. 

\bibliographystyle{ACM-Reference-Format-Journals}
\bibliography{tosem}

\end{document}